\documentclass[12pt]{article}
\usepackage{bm}
\usepackage{color}
\pdfoutput=1
\usepackage{jheppub}
\usepackage{amsfonts,amsbsy,amsmath,amssymb}

\newcommand{\be}{\begin{equation}}
\newcommand{\ee}{\end{equation}}
\newcommand{\ba}{\begin{array}}
\newcommand{\ea}{\end{array}}
\newcommand{\baa}{\begin{array}}
\newcommand{\eaa}{\end{array}}
\newcommand{\bea}{\begin{eqnarray}}
\newcommand{\eea}{\end{eqnarray}}

\newcommand{\R}{\mathbb{R}}
\newcommand{\Pf}{\mathrm{Pf}}
\title{On the fractional instanton liquid picture of the Yang-Mills
vacuum and Confinement}

\author[a,b]{Antonio Gonz\'alez-Arroyo }

\affiliation[a]{ Instituto de F\'{\i}sica Te\'orica UAM/CSIC,  Nicol\'as
  Cabrera 13-15, \\
  Universidad Aut\'onoma de Madrid, E-28049 Madrid, Spain}
\affiliation[b]{
 Departamento de F\'{\i}sica Te\'orica,  M\'odulo 15,
 \\
  Universidad Aut\'onoma de Madrid, Cantoblanco, E-28049 Madrid, Spain}

\emailAdd{antonio.gonzalez-arroyo@uam.es}

\abstract{ I review the main  features of our model of
the 4-dimensional Yang-Mills theory vacuum as a liquid of fractional
instantons. The model provides a possible microscopic mechanism  for
Confinement in four dimensional Yang-Mills theory at $T=0$. It also
connects this property to other non-perturbative properties of the
theory which can be explained by the same model.
This paper is  a, somewhat enlarged,  written up version of my recent talks on the
subject given at  ICTS-Bangalore, KITP-Santa
Barbara and IMSC-Chennai.

}
\begin{document}
\preprint{%
{\flushright \vskip -0.5cm
\vbox{IFT-UAM/CSIC-23-20}
}}%




\maketitle



\section{The model in a nutshell}
\label{s.nutshell}
In a set of journal papers written in the decade of the
90's~\cite{GonzalezArroyo:1995zy, GonzalezArroyo:1995ex}
our group at Universidad Autonoma de Madrid proposed a model of the
4-dimensional Yang-Mills vacuum which essentially describes it   as a
liquid whose basic building blocks are, locally self-dual (and
anti-self-dual), fractional  topological charge lumps  that can be 
henceforth called fractional instantons (and anti-instantons).
This idea has the power of providing a connection and unified
explanation  of many of the  well-known  properties of 
the Yang-Mills vacuum, such as Confinement and the
existence of a finite topological susceptibility. It also hints  at
explaining the finite temperature phase transition and, with the
addition of non-dynamical quarks, the quark condensate and spontaneous chiral
symmetry breaking. 

We should clarify, first of all, that  fractional instantons cannot
exist isolated, so that their size is dictated by the distance to its
neighbours. This is why we describe our model as a liquid and not a
gas. Furthermore, the typical size-separation of the instanton is
determined by quantum fluctuations around them and is therefore fixed
in terms of the inverse lambda parameter of the theory. For the SU(2)
case we estimated this size-separation  to be 0.7 fermi 
 (a unit  defined by declaring the string tension $\sigma$ to be 5
 fermi$^{-2}$).

It should be understood that, although  fractional instantons
owe its stability to its local self-dual or  anti-self dual character, 
the typical configurations in our model are not  global solutions of the classical
equations of motion, since they contain both fractional instantons and
anti-instantons. Furthermore, the fractional instantons locations are
labels that describe not just a smooth configuration, but also its
aura produced by quantum fluctuations around it and contributing to
its free energy.  However, when analyzing some long distance properties
as Confinement, these short-distance fluctuations are irrelevant and
good estimates can be obtained from the position, sizes and fluxes of
these fractional instantons.  This is of course in
the standard spirit of a semiclassical approximation, and would not
have to be mentioned if not so frequently misunderstood.

The idea emerged, as will be explained later, from the study of the
evolution of the system as a function of the spatial volume. It was
first announced in the Lattice Conference in Dallas and hinted in its
corresponding proceedings publication~\cite{GarciaPerez:1993jw}. 
The proposal, however, appears formulated 
explicitly in section 4  of Ref.~\cite{GonzalezArroyo:1995zy}.  Later with two of my PhD
students, Pablo Martinez and Alvaro Montero, we tried to present
further evidence in a series of papers~\cite{GonzalezArroyo:1995ex,
GonzalezArroyo:1996jp}  and lattice conference
proceedings~\cite{ Arroyo-Gonzalez:1994hxl, GonzalezArroyo:1995fu,  GonzalezArroyo:1996gs}. All this work 
is included in their corresponding  doctoral thesis~\cite{PMartinez,
AMontero}. Driven by computational limitations all our numerical
simulations concentrated basically in SU(2) gauge theory.

Let us summarise briefly the main ingredients of the fractional
instanton liquid  model (FILM)  which, in our
opinion, make it very appealing. In the  rest of the paper we will
explain  in more detail these ingredients as well as the evidence that
we obtained in their support.
\begin{enumerate}
\item {\bf Provides a microscopic mechanism for Confinement in 4D
Yang-Mills field theory at zero temperature} \\
As will be explained later this is based on two main features of
fractional instantons: these objects carry flux ($Z_N$  flux)
and have finite free energy. Flux is essential for the confinement
property (in this respect ordinary instantons are useless). The
finite free energy is shared with vortices in 2D and monopoles in 3D,
which are known to lead to confinement. Our mechanism is thus specific
of 4D Yang-Mills theories as a microscopic mechanism should be.
\item {\bf Fractional instantons are smooth classical configurations that
can be thought as the building blocks of more complex self-dual gauge
field configurations}\\
Specific smooth multi-fractional instanton configurations can be easily
constructed and analysed. To construct them one uses properties of the
topology of gauge fields on the 4-torus: twisted boundary conditions. 
These configurations should be regarded as periodic arrays of
fractional instantons in $\R^4$. In this respect neither the torus nor
the boundary conditions are essential, but just tools used to construct
them. The periodicity can be reduced to 1, 2 or 3 dimensional
crystals. We argue, based  on the index theorem, that the crystal
structure can be deformed at no cost to produce a liquid.
\item{\bf N dependence}\\
All the fractional instantons can be thought as made out of  unit
fractional instantons having topological charge $1/N$. The classical
action of these unit fractional instantons remains finite in the large
$N$ limit. This is crucial since this limit does not destroy
Confinement and other non-perturbative properties of the theory. In
this respect instantons  are always argued not to survive the limit.
Furthermore, the existence of these fractional topological charge
objects is also important to understand the theta dependence of the
free energy of the theory for arbitrary $N$. 
\item{\bf They arise naturally in the semiclassical description of the
theory}\\
This is precisely the way in which we got interested in fractional
instantons and what led ultimately to our proposal. We studied the
theory at finite volume interpolating from the small volume
semiclassical regime to the large volume non-perturbative regime. This will be
described later. The $T_3\times \R$ topology used in our study has the
advantage of being less prone to intermediate size-driven phase
transitions which will spoil the connection. Nonetheless, even if a
different topology as $T_2\times \R^2$ is used, ultimately the same
picture of the Yang-Mills vacuum should arise as the volume gets large. 
Indeed,   our proposal is fully 4D rotational invariant.
\item{\bf Provides a possible explanation of some  other properties of 
 Yang-Mills field theory}\\
Apart from Confinement the Yang-Mills theory is {\em known} to have a finite
topological susceptibility which survives the large $N$ limit. Our
model links this phenomenon to Confinement as resulting from the same
configurations. This contrasts with other explanations. 

Our old papers restricted to the study of pure Yang-Mills theory at
T=0. However, the model also can provide an understanding of the
finite temperature phase transition and of  the condensate of 
non-dynamical fermion fields submerged in the medium. This will be
explained later. 
\end{enumerate}

To conclude, and to eliminate misunderstandings,  let us declare with precision 
what is the scope of our model. It is a model of the vacuum of pure
Yang-Mills theory in four flat euclidean dimensions and infinite
space-time. Thus, it is not a lattice model, although the lattice can
been used to disentangle its existence and properties. It is also not
a model attached to any compact manifold with specific boundary
conditions. Topology, boundary conditions and semiclassical tools are
irrelevant in our final formulation,  although they have played a
fundamental role in reaching these conclusions. 

As mentioned earlier, in the remaining part of this paper we will
explain the previous points in greater detail and we will also
present the evidence that supports them. However, before doing that,
we want to  summarise the work by other authors prior to ours, which
contain elements in common or are precursors of our model.

\section{Previous related work}
The closest model to ours is the proposal made by Callan, Dashen and
Gross~\cite{Callan:1977qs, Callan:1977gz},
which also assumed the dissociation of instantons into
constituents.  The building blocks were taken to be
merons~\cite{deAlfaro:1976qet},
which are singular configurations. At the time when those papers were
written the existence of the fractional instantons was unknown. In
some sense we can consider our proposal a  modification of
that of Callan, Dashen and Gross (CDT) in which merons are replaced by
genuinely smooth self-dual and anti-self dual instanton constituents:
fractional instantons. The way in which one explains how the 
typical size-distance of the liquid is fixed, also works in the same way as in
their model.

Other interesting old papers that have some relation to our proposal
followed the paper by Savvidy~\cite{Matinyan:1976mp, Savvidy:1977as} in which he proposed that the
Yang-Mills vacuum was permeated by a uniform magnetic field. At the
classical level such a configuration is unstable and Nielsen and
Olesen~\cite{Nielsen:1979xu} studied the deformation and concluded that the configuration develops
structures forming   some kind of lattice perpendicular to the direction
of the original field. The most interesting connection comes because
these the individual components do indeed correspond to fixed `t Hooft
flux, just like fractional instantons. 

Another previous attempt worth mentioning is the Instanton liquid
model~\cite{Shuryak:1981ff, Shuryak:1993kg}. This is actually similar in describing a dense media of
topological structures. However, the main problem is that the basic 
components are taken to be instantons. Apart from being incapable of
describing confinement, since instantons carry no flux, this model has
a serious difficulty in dealing with the large $N$ limit. The action
of an instanton and hence its  free energy diverges exponentially with
$N$ in that limit. A nice feature of fractional instantons is that
indeed their action remains finite in that limit. Our model is thus a
modification of the liquid model involving a fractionalization of the
instantons as in the CDT proposal but with the basic building blocks
being genuinely local topological structures. Still some of the
properties of the instanton liquid model concerning non-flux related
observables might still hold.

Finally, it is also worth mentioning the work of
Zhitnitsky~\cite{Zhitnitsky:1989ds, Zhitnitsky:1991eg}
who advocated the presence of fractional topological charge objects to
explain the dependence of the free energy on $N$ in the presence of a
non-zero  $\theta$ parameter. 

Curiously our proposal combines elements of all the previous ones
solving some of the problems that they had.

\section*{On Confinement}
One of the important implications    of our proposal  was that the
fractional-instanton-liquid model (FILM) can explain Confinement. This property
was originally  proposed as an explanation of the impossibility of 
producing free quarks in high energy Physics collisions of hadrons. 
The work done by many authors during the 70 and 80 decades led to an
elucidation of the nature of this phenomenon. Sticking to SU(N)
Yang-Mills theory makes things simpler: the potential among two
(non-dynamical) sources in representations which are not blind to 
the center of the group (like the fundamental) grows linearly with the separation. This leads 
to an operational criterion for Confinement as formulated  by
Wilson~\cite{Wilson:1974sk}:
the area law of large Wilson loop expectation values. The slope of the linear
dependence of the potential energy with distance is the string tension
$\sigma$, which acts as an order parameter for Confinement. 
This led immediately to examples of statistical mechanical models
which have confining and non-confining phases. The simplest example of
a Confining system for all values of the coupling is 2-dimensional  $Z_N$ gauge theories,  in
which one can easily do a back-of-the-envelope calculation  of the
string tension. The main reason is that it costs a finite amount of free-energy to introduce a
plaquette carrying non-zero flux (a discrete vortex) within the
interior of the Wilson loop. 

In 3-dimensions the conservation of flux  (Bianchi identity) imposes that
flux appears in the form of flux tubes, which have to be arbitrarily
large in order to produce Confinement. These systems do not show
Confinement at weak coupling since  both energy and entropy depend
linearly on the length and at sufficiently small coupling energy
always wins. The corresponding calculation is essentially identical
with the Peierls contour ones for spin systems in 1 dimension less. 
Thus, in 3 dimensions one needs sources of flux: monopoles. These monopoles
are point-like in three dimensions and have a finite free energy. 
Proof of Confinement in 3D theories  with monopoles 
have been given, both for simple statistical models as for continuum
counterparts. A notable  example of the  latter is  the proof given by  Polyakov 
that three dimensional abelian gauge
theories with monopoles confine~\cite{Polyakov:1975rs}.

However,  in 4 dimensions monopoles develop also into world-lines,  which
bring back the same problem observed for vortices when moving from 2 to
3 dimensions. Thus, abelian gauge theories with monopoles do not
confine for all values of the coupling in 4D. 

Summarizing, it is clear from the
examples of statistical mechanical models that Confinement is a natural 
phenomenon for gauge theories,  but depending on the values of the parameters 
and the dimension of space-time it might hold or not. 

If we now look at the continuum field description one finds a very
nice interpretation of the nature of Confinement. A simple idea that
was pointed out by Parisi is that if magnetic monopoles
existed, the potential energy  between a monopole and
antimonopole in a superconductor would grow linearly with distance. This gives a nice
physical explanation for the linear dependence: the flux created by a
quark source travels in 1 dimensional tubes rather than spreading
isotropically over space. This picture was applied to gauge theories and 
came to be known as the Dual Superconductor
mechanism~\cite{tHooft:1975krp, Mandelstam:1974pi}.

The question then is whether four-dimensional Yang-Mills theory is
indeed confining. Over the last 40 years convincing evidence that this
is the case has been provided by the Lattice Gauge Theory community
starting with the pioneering work of Mike Creutz~\cite{Creutz:1980zw}. Furthermore,
the value of the string tension has been determined in terms of other
units. Obviously, this is a numerical proof and not a mathematical one. 
Indeed, testing Confinement  involves taking two limits: infinitely
large Wilson loops and infinitely small bare couplings. None of the
two can be achieved numerically.  To see what the situation is I will make use
of my own computation of a quantity which I consider very interesting:
\be
\tilde{F}(r,\tau) = -\frac{\partial^2 \log W(r',t')}{\partial r' \partial
t'}\left|_{r'=r\sqrt{\tau}, t'=t/\sqrt{\tau}}\right. 
\ee
where $W(r,t)$ is the expectation value of a rectangular Wilson loop
of size $r\times t$. In contrast to $W$ itself, $\tilde{F}$ is
ultraviolet finite and depends on a linear parameter $r$ and the
aspect ratio $\tau$. We can take for  definiteness $\tau=1$ and study
the behaviour of this observable  as a function of the linear size and
the bare `t Hooft coupling constant $\lambda_L$. The result obtained
in Ref.~\cite{GonzalezArroyo:2012fx} is displayed in
Fig.~1 as a function of
the inverse size of the loop square. Confinement means that the function
extrapolates to a non-zero value as $1/r^2\longrightarrow 0$. The reason
why we believe this is true is not just because it seems so by
prolonging the data by eye. The fact  is that the approach is linear
in $1/r^2$ with a slope consistent with $\pi/12$. This is the so-called
Luscher term~\cite{Luscher:1980fr}  which is the expected behaviour for the
flux-tube model. Extrapolation is hence analytically under control. At the same
time  the curve appears to be the same for all values of the bare coupling. 
Thus,  although the bare coupling is not too small, all the dependence
is absorbed in the value of $\bar{r}(\lambda_L)$ defined  by an ad hoc
condition $\bar{r}^2 \tilde{F}(\bar{r},\tau)=1.65$. This a
manifestation of scaling,  and indeed the  dependence on the scale on
the coupling follows approximately the prediction given by the
perturbative beta function of the theory. Hence, also the so-called
continuum limit corresponding to $\lambda_L\longrightarrow 0$ goes
according to theoretical expectations.

\begin{figure}
\begingroup
  \makeatletter
  \providecommand\color[2][]{%
    \GenericError{(gnuplot) \space\space\space\@spaces}{%
      Package color not loaded in conjunction with
      terminal option `colourtext'%
    }{See the gnuplot documentation for explanation.%
    }{Either use 'blacktext' in gnuplot or load the package
      color.sty in LaTeX.}%
    \renewcommand\color[2][]{}%
  }%
  \providecommand\includegraphics[2][]{%
    \GenericError{(gnuplot) \space\space\space\@spaces}{%
      Package graphicx or graphics not loaded%
    }{See the gnuplot documentation for explanation.%
    }{The gnuplot epslatex terminal needs graphicx.sty or graphics.sty.}%
    \renewcommand\includegraphics[2][]{}%
  }%
  \providecommand\rotatebox[2]{#2}%
  \@ifundefined{ifGPcolor}{%
    \newif\ifGPcolor
    \GPcolortrue
  }{}%
  \@ifundefined{ifGPblacktext}{%
    \newif\ifGPblacktext
    \GPblacktexttrue
  }{}%
  \let\gplgaddtomacro\g@addto@macro
  \gdef\gplbacktext{}%
  \gdef\gplfronttext{}%
  \makeatother
  \ifGPblacktext
    \def\colorrgb#1{}%
    \def\colorgray#1{}%
  \else
    \ifGPcolor
      \def\colorrgb#1{\color[rgb]{#1}}%
      \def\colorgray#1{\color[gray]{#1}}%
      \expandafter\def\csname LTw\endcsname{\color{white}}%
      \expandafter\def\csname LTb\endcsname{\color{black}}%
      \expandafter\def\csname LTa\endcsname{\color{black}}%
      \expandafter\def\csname LT0\endcsname{\color[rgb]{1,0,0}}%
      \expandafter\def\csname LT1\endcsname{\color[rgb]{0,1,0}}%
      \expandafter\def\csname LT2\endcsname{\color[rgb]{0,0,1}}%
      \expandafter\def\csname LT3\endcsname{\color[rgb]{1,0,1}}%
      \expandafter\def\csname LT4\endcsname{\color[rgb]{0,1,1}}%
      \expandafter\def\csname LT5\endcsname{\color[rgb]{1,1,0}}%
      \expandafter\def\csname LT6\endcsname{\color[rgb]{0,0,0}}%
      \expandafter\def\csname LT7\endcsname{\color[rgb]{1,0.3,0}}%
      \expandafter\def\csname LT8\endcsname{\color[rgb]{0.5,0.5,0.5}}%
    \else
      \def\colorrgb#1{\color{black}}%
      \def\colorgray#1{\color[gray]{#1}}%
      \expandafter\def\csname LTw\endcsname{\color{white}}%
      \expandafter\def\csname LTb\endcsname{\color{black}}%
      \expandafter\def\csname LTa\endcsname{\color{black}}%
      \expandafter\def\csname LT0\endcsname{\color{black}}%
      \expandafter\def\csname LT1\endcsname{\color{black}}%
      \expandafter\def\csname LT2\endcsname{\color{black}}%
      \expandafter\def\csname LT3\endcsname{\color{black}}%
      \expandafter\def\csname LT4\endcsname{\color{black}}%
      \expandafter\def\csname LT5\endcsname{\color{black}}%
      \expandafter\def\csname LT6\endcsname{\color{black}}%
      \expandafter\def\csname LT7\endcsname{\color{black}}%
      \expandafter\def\csname LT8\endcsname{\color{black}}%
    \fi
  \fi
  \setlength{\unitlength}{0.0500bp}%
  \begin{picture}(8220.00,4818.00)%
    \gplgaddtomacro\gplbacktext{%
      \csname LTb\endcsname%
      \put(946,704){\makebox(0,0)[r]{\strut{} 1}}%
      \put(946,1254){\makebox(0,0)[r]{\strut{} 1.5}}%
      \put(946,1804){\makebox(0,0)[r]{\strut{} 2}}%
      \put(946,2354){\makebox(0,0)[r]{\strut{} 2.5}}%
      \put(946,2903){\makebox(0,0)[r]{\strut{} 3}}%
      \put(946,3453){\makebox(0,0)[r]{\strut{} 3.5}}%
      \put(946,4003){\makebox(0,0)[r]{\strut{} 4}}%
      \put(946,4553){\makebox(0,0)[r]{\strut{} 4.5}}%
      \put(1078,484){\makebox(0,0){\strut{} 0}}%
      \put(2304,484){\makebox(0,0){\strut{} 2}}%
      \put(3531,484){\makebox(0,0){\strut{} 4}}%
      \put(4757,484){\makebox(0,0){\strut{} 6}}%
      \put(5983,484){\makebox(0,0){\strut{} 8}}%
      \put(7210,484){\makebox(0,0){\strut{} 10}}%
      \put(176,2628){\rotatebox{-270}{\makebox(0,0){\strut{}$\bar{r}^2
      \tilde{F}(r,1)$  }}}%
      \put(4450,154){\makebox(0,0){\strut{}$2 \frac{\bar{r}^2}{r^2}$}}%
    }%
    \gplgaddtomacro\gplfronttext{%
      \csname LTb\endcsname%
      \put(2926,4380){\makebox(0,0)[r]{\strut{}$\lambda_L=2.80179$}}%
      \csname LTb\endcsname%
      \put(2926,4160){\makebox(0,0)[r]{\strut{}$\lambda_L=2.74973$}}%
      \csname LTb\endcsname%
      \put(2926,3940){\makebox(0,0)[r]{\strut{}$\lambda_L=2.72869$}}%
      \csname LTb\endcsname%
      \put(2926,3720){\makebox(0,0)[r]{\strut{}$\lambda_L=2.68902$}}%
      \csname LTb\endcsname%
      \put(2926,3500){\makebox(0,0)[r]{\strut{}$\lambda_L=2.62134$}}%
      \csname LTb\endcsname%
      \put(2926,3280){\makebox(0,0)[r]{\strut{}Linear part}}%
      \csname LTb\endcsname%
      \put(2926,3060){\makebox(0,0)[r]{\strut{}Quadratic fit}}%
    }%
    \gplbacktext
    \put(0,0){\includegraphics{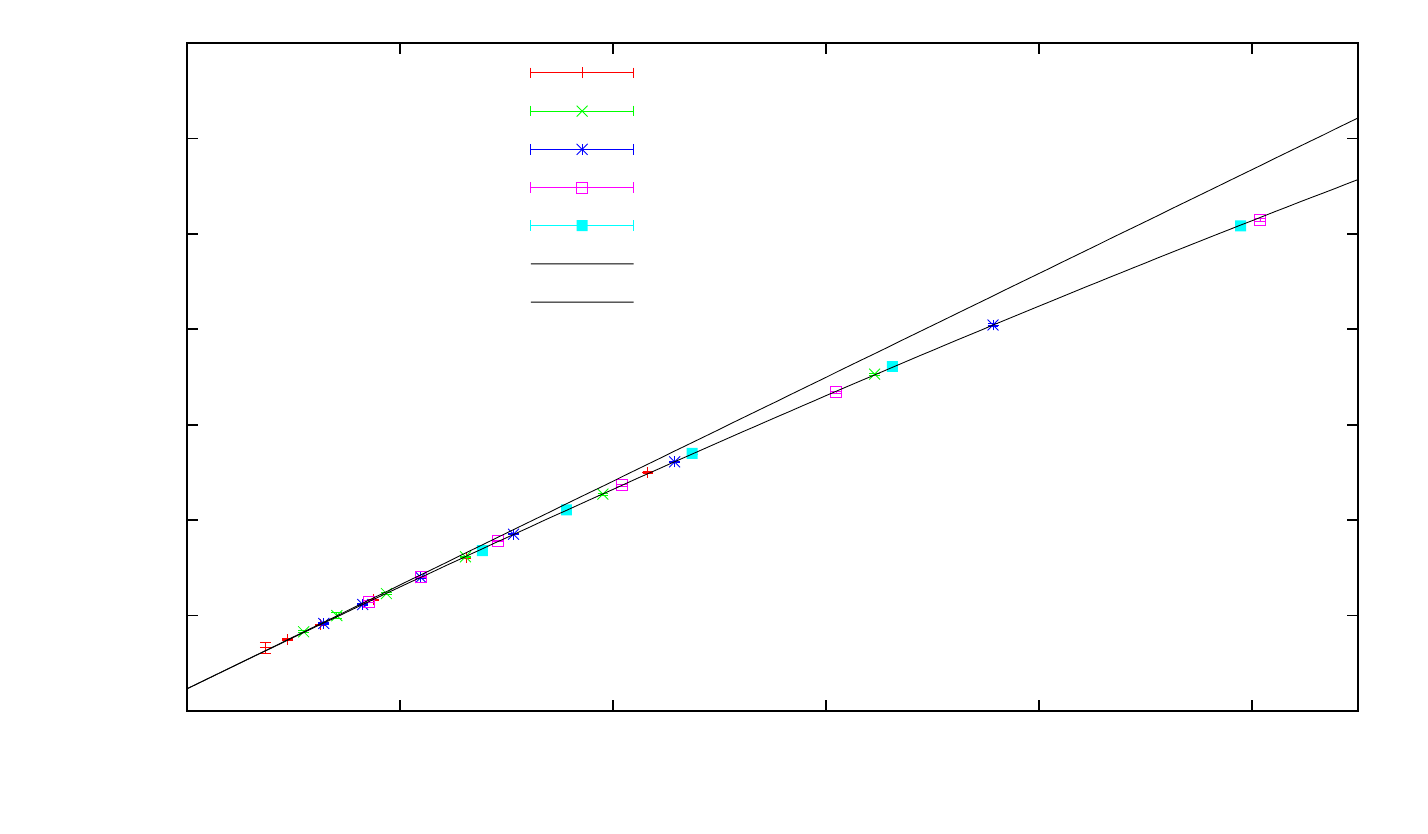}}%
    \gplfronttext
  \end{picture}%
\endgroup
\label{fig:Ftilde}
\caption{The continuum function $\bar{r}^2 \tilde{F}(r,1)$ for SU(8)
Yang-Mills in units of $ \bar{r}$ defined by the condition $\bar{r}^2
\tilde{F}(\bar{r},1)=1.65$}
\end{figure}

In conclusion, in my opinion,  the evidence for Confinement in 4D
Yang-Mills theory is completely convincing from a physicist perspective.
Is this all? I do not believe that is the end of the story.  We still 
lack an  understanding of the microscopic mechanism driving Confinement 
in these theories. One should not
confuse the explanation of the nature of the phenomenon which is universal and
understood several decades ago with the microscopic mechanism. To
understand this important distinction one should consider the analogy
with another phenomenon of Nature: Superconductivity. The understanding
of the origin of this phenomenon is universal: the condensation of a charged
field.  This was known since the work of the London
brothers~\cite{London}. 
A simple relativistic model describing the phenomenon a la
Ginzburg-Landau is the Abelian Higgs model, involving a charged scalar
field with a Mexican hat potential. Nonetheless, the
microscopic theory corresponding to ordinary superconductor materials
should explain where does this effective scalar field come from and
what mechanism drives its condensation. This is provided by the BCS
theory~\cite{Bardeen:1957kj,Bardeen:1957mv}. However, this microscopic mechanism is system-dependent 
and might be different  for other systems
like high $T_C$ superconductors or others.

In much the same way,  the microscopic mechanism underlying
Confinement in pure Yang-Mills theory in 4 euclidean dimensions is yet
to be fully understood. Our fractional instanton liquid model aims at
producing such an explanation. As mentioned in the first section of
this paper there are several properties  that fractional instantons
have that make it a good candidate: they carry flux, have finite
action density, survive the large N limit, etc . Can we provide some
evidence?  Relating the expectation values
of Wilson loops with bulk properties is hard as it deals with the
non-abelian Stokes theorem. In an abelian setting, Stokes theorem
allows to connect the Wilson loop expectation value with the flux that
traverses a surface with edge at the Wilson loop. Fortunately, our
investigation of fractional instantons showed that the small Wilson
loops surrounding the fractional instantons are  approximately given
by elements of the center of the SU(N) gauge group. This allows using 
abelian arguments to explain why a liquid of independent fractional
instantons does indeed produce area law i.e. Confinement. This is the
argument given in Ref.~\cite{GonzalezArroyo:1995zy}.  This also gives a rough connection
between the string tension and the topological susceptibility. Notice,
that for Confinement to hold, the fractional instantons should behave
as independent items. Meson-like instanton-antiinstanton bound pairs will
not give area law. The same applies for ordinary instantons which
appear as baryons containing N fractionals and hence neutralizing the
total flux. Dissociation is a must.

Our arguments give a reasonable qualitative explanation of why the
liquid model confines. In trying  to give an rigorous  proof  one 
has to face the difficulty of not having an analytic expression of these fractional instantons
except in some special cases. Furthermore, even if one had it, the
non-linearity of the Yang-Mills field equations makes it is very difficult 
to create a reliable ansatz for a  multi-instanton anti-instanton  configuration on
which to measure the Wilson loop. It is for this reason that together
with my students we  attempted to give a numerical proof.  For that
purpose,  we created configurations having many fractional instantons
at random locations and measured the Wilson loop on them. The
resulting value  was seen to approach the area law for large sizes.
This result appeared in a paper with the
self-explanatory title of {\em Do classical configurations produce
Confinement?}~\cite{GonzalezArroyo:1996jp}.
To produce  these configurations we started by  creating a configuration 
consisting of a  crystal  of fractional instantons. This is easy to do,
as will be explained later. Then we 
modified this configuration by a series of heating and cooling steps. 
The goal of this process is  to disorder the  position of the
fractional instantons. The reason why this works is because, as will
be mentioned later, the position of the fractional instantons is part
of the moduli and therefore disordering the crystal costs zero action. 
The  heating steps, if moderate, modify  the positions but also introduce
short distance noise, which the subsequent cooling eliminates. 
Obviously, our random configurations do not reflect a typical vacuum
one  since they are all made of fractional instantons  of the same charge.
However, from the point of view of flux, the instantons and
anti-instantons behave similarly, so we expect the property to hold
also if both fractional instantons and anti-instantons are present. 
Indeed, Monte Carlo generated configurations seem to show the same
type of relation between the density of structures and the string
tension. 

\section{On fractional instantons}
The work of Polyakov~\cite{Polyakov:1975rs} put forward the relevance of  solutions
of the classical equations of motion in euclidean space in the path integral formulation
of quantum field theories. At the same time the BPST instanton
solution~\cite{Belavin:1975fg} was presented. Its stability is related to
topological considerations. Mathematically the  topology of gauge
fields is better understood when formulated on a compact manifold. For
U(N) bundles we have Chern classes characterizing  inequivalent
bundles. The inequivalence translates to the corresponding connections. 
In 4 dimensions the second Chern number, instanton number or
topological charge $Q$ separates gauge fields into sectors. The BPST
instanton is a classical solution in euclidean space belonging to the
Q=1 sector. Although, space-time is non-compact the topological
notions apply by imposing finite action. Topologically the bundle is
equivalent to one formulated in $S_4$. 

After the instanton was found a fascinating quest started to find the
corresponding solutions for all values of $Q$. This led to the ADHM
construction~\cite{Atiyah:1978ri} giving the recipe to construct all multi-instanton solutions 
in $S_4$ (or infinite space-time with finite action)   in terms of the solution 
of an algebraic problem. However, this does not provide an ansatz for the
structure of such a general solution. There are of course some
approximate solutions that describe a collection of instantons with
separations much larger than their sizes. This can be described as a
dilute gas of instantons. Unfortunately, this has led many scientists
to believe that a general self-dual configuration can be simply
described as a collection of instantons. Explicit examples show that
this is not the case. 

When thinking about configurations that populate the Yang-Mills vacuum
the condition of finite action is obviously wrong. The action is an
extensive quantity and physically one can only demand finite action
density. Furthermore the picture of a dilute instanton gas is also
very inappropriate. This envisages that space is mostly covered  by the
classical vacuum (a pure gauge=flat connection) up to islands where the instantons
live. We rather expect the vacuum to be a very dense medium mostly far
from the classical vacuum almost everywhere. There is another reason
to expect this: quantum fluctuations. The calculation of quantum
fluctuations around a BPST instanton performed by `t
Hooft~\cite{tHooft:1976snw}
showed that these fluctuations break the dilatation invariance of the
classical theory. This makes that instantons of larger size  are more
probable than those of smaller size. This points against diluteness. 

A new ingredient appeared when `t Hooft realized that in an SU(N)
theory on the torus the topology of the bundle allows for new sectors
characterized by $Z_N$ flux transversing each of the faces of the
torus~\cite{tHooft:1979rtg, tHooft:1980kjq, tHooft:1981sps}~\footnote
{These fluxes are the remnants of
first Chern numbers when projecting the bundle from U(N) down to SU(N).
In Mathematics these are associated to the so-called Stiefel-Whitney
characteristic classes and provide an obstruction to lifting the
bundle from SU(N)/Z$_N$ back to U(N). The reader who is not familiar with
these ideas can consult my rather simple introduction for a Physics
audience~\cite{GonzalezArroyo:1997uj}.}.
In this case the inequivalence of  bundles does not necessarily
imply inequivalent  connections. This is well-known by mathematicians
but came somewhat as a surprise when zero-action configurations with
twist, often referred as twist-eaters,
appeared~\cite{Groeneveld:1980zx,Ambjorn:1980sm}. 
However, `tHooft also showed that in some twist sectors the topological charge is
non-zero and even given by an integer multiple of $1/N$. If the flux
through the $\mu-\nu$ plane is given by $\exp\{2\pi i n_{\mu
\nu}/N\}\in Z_N$, then the twist is specified by the twist tensor $n_{\mu \nu}$, an
antisymmetric matrix of integers mod $N$. The corresponding
topological charge is given by 
\be
Q=p-\frac{\Pf(n)}{N}
\ee
where $p$ is an integer and $\Pf(n)$ stands for the pfaffian of the
antisymmetric tensor. Hence, certain twist values imply non-zero topological charge 
which implies non-trivial gauge fields. Self-dual (anti-self-dual)  solutions having 
fractional topological charge are referred as fractional instantons (anti-instantons). These
configurations exist and are non-singular. Mathematicians have proven
their existence in various cases~\cite{Sedlacek:1982cd} and an explicit solution was
given by `t Hooft himself~\cite{tHooft:1981nnx}. The solution he found is rather special
since the action density is uniform and it is only self-dual when the
torus sizes have particular ratios. Anyhow, since these solutions are
the minima of the action in each twist sector, it is extremely easy to
obtain the solutions numerically by gradient flow (similar to what
some mathematicians use to prove the existence). Together with my
former student Margarita García Pérez (and Bo S\"oderberg) we applied this procedure to
obtain these numerical solutions for very different sizes  and even
showed convergence when one of the torus sizes tends to infinity
compared to the others. This was part of her doctoral thesis and
appears in the corresponding publications~\cite{GarciaPerez:1989gt,
GarciaPerez:1992fj}. The work was mostly
restricted to SU(2) for computational reasons,  but later the SU(N)
case was studied in a similar way as part of the thesis of Alvaro
Montero~\cite{GarciaPerez:1997fq, AMontero, Montero:2000mv}. In all these cases the fractional instanton appear as
lumps in the action density field having a particular center in
space-time. In Fig.~2 we show the two-dimensional action
density profile  in a spatial plane at its center of a $Q=1/2$ fractional instanton for SU(2).
When talking about the action for classical solutions it should be
understood that the `t Hooft coupling is fixed to 1. With this
convention the action of an instanton equals $8 \pi^2$.

\begin{figure}
\includegraphics[width=\linewidth]{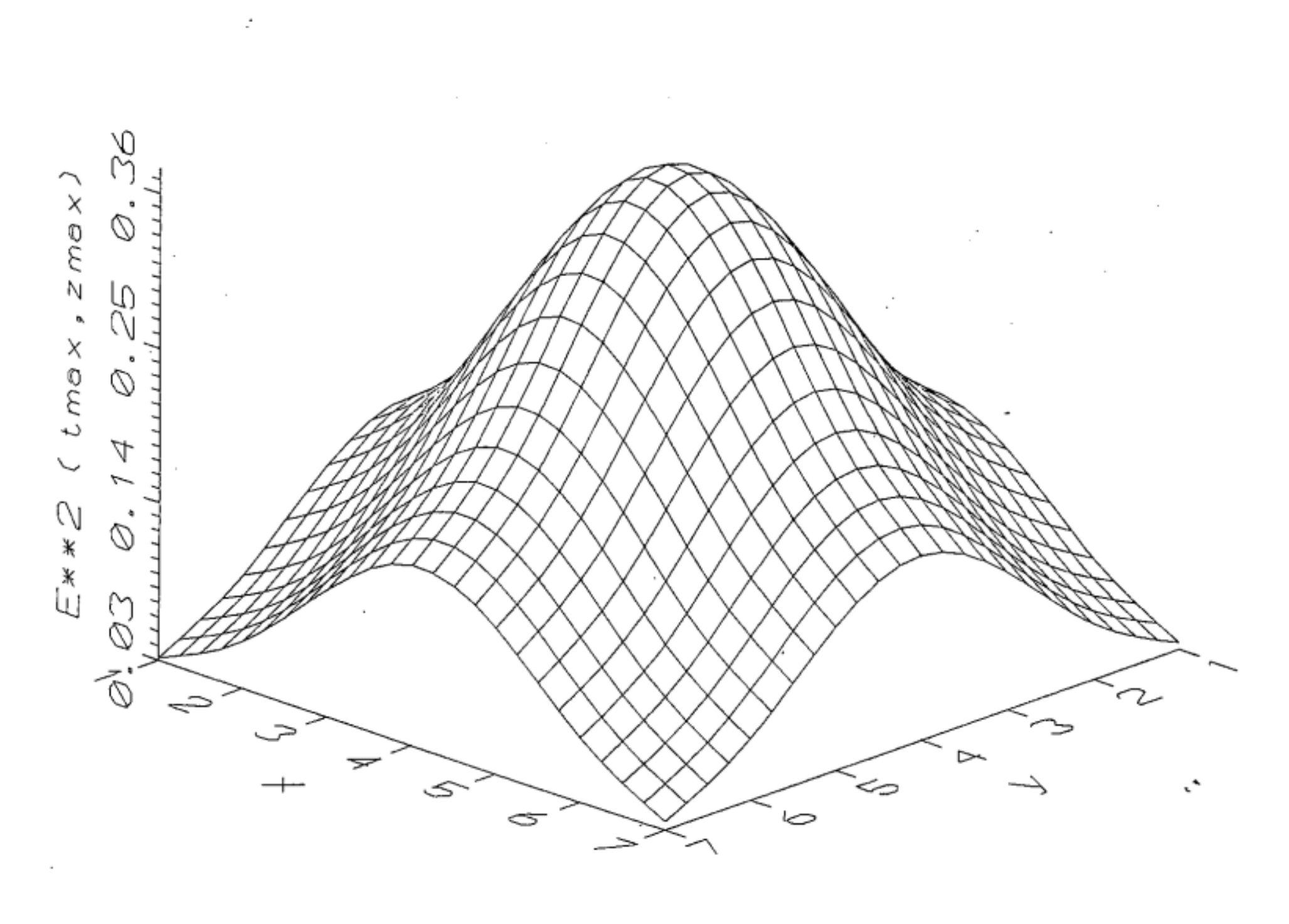}%

\label{fig:lump}
\caption{This figure shows the action density profile for an SU(2)
Q=1/2 fractional instantons as a function of
two spatial coordinates and fixing the other spatial and time
coordinates to the maximum.}
\end{figure}

A very important breakthrough appeared when Kraan-Van Baal and Lee-Lu
discovered non-trivial holonomy  calorons~\cite{Lee:1998vu,
Lee:1998bb,Kraan:1998sn,Kraan:1998pm}. These are analytic
self-dual Q=1 solutions  living in $S_1\times \R^3$. The action density
profile of these solutions show that in certain regions of parameter
space the solution is made out of $N$ lumps, each carrying a fraction
of the topological charge (See Fig.~3 for an example for
SU(3)).   When the holonomy is invariant under center
symmetry all the lumps have equal $1/N$ topological charge. These
lumps were called constituent monopoles by Pierre van Baal, but some
authors refer to them as instanton-monopoles. 
Curiously,  in other regions of parameter space, the configuration
looks just like a single Q=1 instanton lump. The transition between
these situations  is continuous. One can regard the constituent monopoles
as examples of fractional instanton solutions and the process shows
that they can fuse to produce ordinary instantons. Thus, it makes
sense to consider  fractional instantons as constituents of ordinary
instantons. The connection was made clearer when it was shown  that
calorons appear as limits of the standard solutions on the 4-torus and
that the individual constituents can be separated out by imposing
twisted boundary conditions~\cite{GarciaPerez:1999hs}.

h
\begin{figure}
\includegraphics[width=\linewidth]{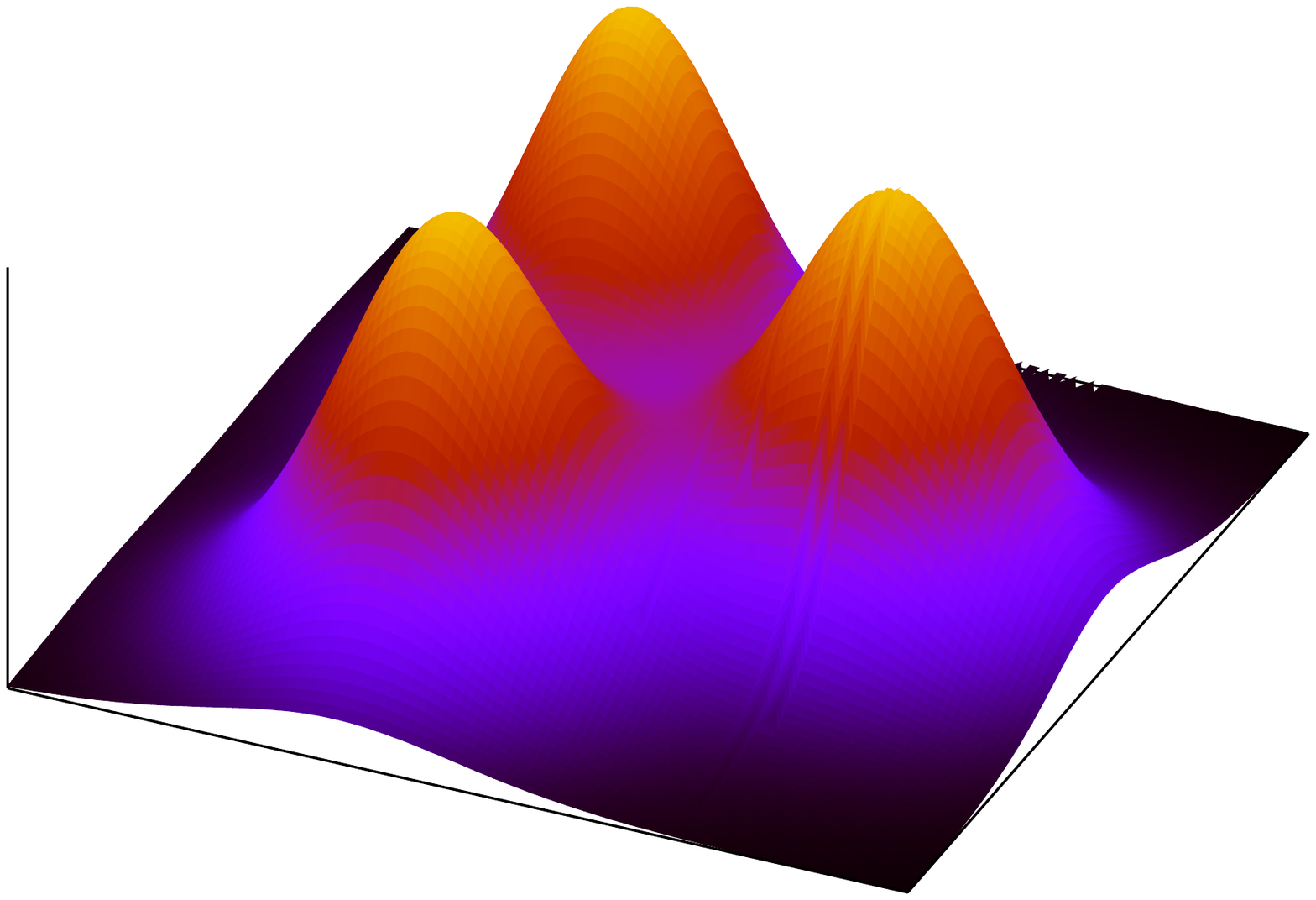}%
\label{fig3}
\caption{This figure shows the action density profile in one spatial
plane for one $Q=1$ caloron solution for SU(3). The profile shows how
the total action ($8 \pi^2$) and topological charge is split into 3
substructures of fractional charge.}
\end{figure}

At the classical level different shapes and appearances depend on the
choices of the torus sizes and the twist tensor (fluxes through the
faces). For example if two of the torus periods are much larger than
the other two one can find solutions that tend asymptotically to
solutions in $T_2\times \R^2$~\cite{GonzalezArroyo:1998ez,
Montero:1999by, Montero:2000pb}. We labelled these solutions vortex-like
since in the $\R^2$ plane they look pretty much like the
Abrikosov-Nielsen-Olesen vortices of the abelian Higgs model. They are
also exponentially localized. Of course the difference is that the
size is not dictated by the parameters of the lagrangian but by the
size of the small torus. More importantly Wilson loops around these
objects have non vanishing $Z_N$ flux. In Fig.~4 we display
the action density profile of one of these solutions (as an $\R^4$
configuration) as seen in each
of the two-dimensional sections. In Mathematical literature these kind
of solutions are often referred to as {\em doubly-periodic instantons}.

\begin{figure}
\includegraphics[width=\linewidth]{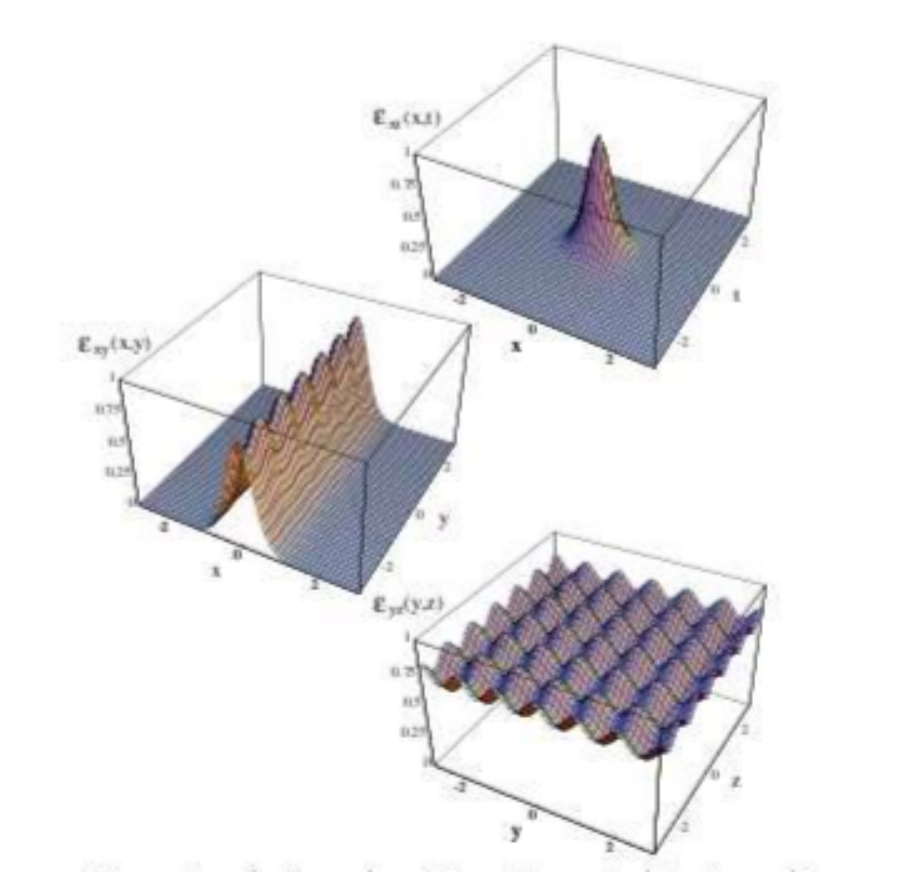}%
\label{fig4}
\caption{This figure shows action density profile of the self-dual
vortex-like solution for SU(3) depicted as an $\R^4$ classical
solution. Depending on the plane of projection the solution looks like a localized
structure, a one dimensional periodic array, or a two-dimensional array. }
\end{figure}

In summary, fractional instantons exist as smooth solutions of the
Yang-Mills euclidean classical equations. A unit fractional instanton
with topological charge $1/N$ can be constructed by considering a torus 
with appropriate twisted boundary conditions and minimizing the
action in this bundle.  However, it is important to look
at these objects as solutions in $\R^4$ having periodicities in
space-time. They satisfy the classical equations of motion everywhere,
but have infinite action (but finite action density). 
It has also been shown that solutions exist when some of
the periods go to infinity and the the manifold becomes $T_n\times
\R^{(4-n)}$. This has been seen for $n=3$ (our original small spatial
volume instantons), $n=2$ (the vortex-like solutions just described) 
and $n=1$ (the constituent monopoles in calorons). However, as we will 
argue in the next section neither twist nor the periodicity are
essential for their existence.

\section{Multi-fractional instanton solutions and its moduli}
In the previous section we saw how fractional instantons can be
constructed by considering classical solutions on a torus with well
chosen twisted boundary conditions. Automatically these are also
solutions in a bigger torus containing several periods of the small
torus. The transition functions can be trivially constructed. The
action, the topological charge, the number of lumps  and the flux are
additive so they will be integer multiples of the ones in the small
torus. Indeed the `t Hooft flux is additive modulo  $N$,  so that the
total flux  on the big torus can be made zero. Hence, these
multifractional instantons exist as solutions in a torus without twist. 
Of course, in that case  the topological charge in this big torus 
will  be an integer, and the number of unit fractional  instantons will
be proportional to N. This global constraint is similar to the
situation in a one-dimensional scalar field theory with a double well potential in which kinks  are
solutions with antiperiodic boundary conditions, but even in the
presence of periodic boundary conditions there are solutions with an
even number of kinks.    

The question now is what is the moduli space of the space of self-dual
solutions to which these multi-instanton solutions belong. One can
easily show that  self-dual deformations satisfy an equation: The
Dirac equation in the adjoint representation in the background of the
original solution. This allows the index theorem to be used to
investigate the number of independent deformations as done originally
for solutions on the sphere~\cite{Schwarz:1977az, Atiyah:1977bu}. The result in the
torus case~\cite{AtiyahD2} is that the number is equal to $4QN$: the dimensionality
of the moduli space. Out of these parameters 4 are associated to a symmetry:
space-time translations. The corresponding solution of the adjoint
Dirac equation is the supersymmetric partner of the gauge field (we
will call it the Supersymmetric zero-mode SZM). 
Notice then that if $Q=1/N$ the solution is essentially unique up to
translations. However, for the larger $Q$ case, the remaining degrees of freedom are not
associated to symmetries. Our proposal is that these parameters can be
interpreted as the 4-dimensional locations  of each of the unit
fractional instantons making up the solution. Notice that $QN$ counts
the number of lumps. The moduli space can then be seen as the degrees
of freedom of a 4-dimensional gas (or liquid) with $QN$ particles. 
This is a certain generalization of the situation appearing in the
2-dimensional abelian Higgs model on the torus with flux $q$ (first
Chern number) in which the moduli space is $2q$-dimensional and 
corresponds to a 2-dimensional gas of $q$ vortices~\cite{taubes2}. This
picture applies for the caloron case for which these deformations and
its corresponding adjoint zero-modes have
been computed~\cite{GarciaPerez:2006rt, GarciaPerez:2008gw,
GarciaPerez:2009mg}. Showing that positions can be associated to
moduli parameters can be particularly easy for small deformations.
However we lack a suitable parameterization of the moduli space for
large deformations. Notice that when fractional instantons approach
each other one can cease to see as many lumps as the counting $QN$
indicates. This also happens of course for vortices in the abelian
Higgs model. However, in that case there is a well-defined definition
of the vortex position: the zero of the corresponding Higgs field. It
turns out that the number of such zeroes (counted with multiplicity) is
equal to $q$ (the flux or first Chern number, or number of unit vortices).  
It would be very nice to have an equivalent definition of fractional
instanton position in our case. 

Going back to the general picture, we do interpret all the self-dual
solutions as a space of unit fractional instantons with random
locations in 4-dimensional euclidean space-time. Deformation of the
position of the individual unit fractional instantons do not cost any
action since they are all self-dual and the action saturates  the 
Bogomolny bound. This is a  very 
remarkable fact given the non-linear character of the classical
equations of motion. Of course, that also happens for ordinary
multi-instanton solutions on $S_4$ or multivortices.

\section{Quantum fluctuations around fractional instantons}
The previous section is purely classical but quantum fluctuations are
essential in any description of the theory. It modifies the weight of
a particular configuration within the path integral according to its
free energy. Part of this free energy is given by the classical action
of the solution, but this is not all and there is an additional
contribution of what we previously called the aura of the configuration. 
These fluctuations can be computed at weak coupling using perturbation 
theory. For an ordinary instanton this was done in the remarkable
paper by `t Hooft~\cite{tHooft:1976snw}. One important conclusion is that
fluctuations break the scale invariance of the theory. The free-energy
of instantons depends on the size, making larger instantons more
probable that small ones. This breaking is part of the general feature
of dimensional transmutation and the scale breaking is connected to
the beta function. The renormalization group implies that this quantum
weight can be absorbed into the scale-dependent coupling constant
entering the classical action weight. For fractional instantons a
similar phenomenon is expected so that for small sizes one expects
also a growth of the distance-size of the liquid due to quantum
fluctuations. 

In multi-instanton (fractional or not) cases it would be possible to
compute analytically in perturbation theory the free energy dependence on the 
remaining moduli parameters. For the fractional instanton liquid  this
would suggest whether fractional instantons develop a pattern as 1
dimensional strings like in the spaghetti vacuum, or two dimensional
sheets of vortex-like world-sheets, or more isotropic situations.
For the caloron case there are calculations which represent a step in
this direction~\cite{Diakonov:2004jn, Diakonov:2005qa, Gromov:2005hv,
Gromov:2005ij, KorthalsAltes:2014dkx}.

Returning to the overall scale of our liquid, we proposed that in the Yang-Mills
vacuum the typical size-distance scale is fixed to a certain value in
units of the lambda parameter. What determines the optimal scale is
the balance between two phenomena: having larger fractional instantons
and having a higher density of them. By extrapolating the quantum
weight to larger sizes one reaches a point at which it is more
probable to have another instanton than not to have it. On the basis 
of these semiclassical estimates in Ref.~\cite{GonzalezArroyo:1995zy} we
explained how the average size-distance follows a probability
distribution that has a maximum slightly above the point at which the 
semiclassical formula predicts the breakdown. From our  SU(2) data we
determined this typical size to be $\bar{l}$=0.7 fermi. 
In Ref.~\cite{GonzalezArroyo:1995zy} we show how this number gives us  
a  string tension and  topological susceptibility which match roughly 
with  independent measurements of these quantities. This also explains  
why certain quantities do not change much with size beyond this point. 
Of course, our explanation is similar to the one given by Callan, Dashen
and Gross~\cite{Callan:1977qs, Callan:1977gz}.

\section{The roadmap towards the proposal}
The ideas written in the previous sections do conform our picture of
the Yang-Mills vacuum. This did not come out of the blue but, 
as mentioned earlier,  the proposal resulted from our study of
Yang-Mills fields dynamics as a function of the spatial volume. The
main idea is that for small volumes asymptotic freedom makes
perturbative and semiclassical computations good approximations. As
the volume gets larger the dynamics becomes increasingly strongly
coupled and at some stage the finite volume effects    become
very small and almost irrelevant. This occurs when the spatial linear size is much larger
than the characteristic correlation length of the theory given by the
inverse mass gap or the inverse $\Lambda$ parameter. Much before this
happens the analytic semiclassical techniques are unable to give an
accurate description of the physical observables. Nonetheless,  the
hope is that by monitoring how this transition from small to large
volumes takes place, one can extract information about the
non-perturbative structure and properties of theory.
The starting idea of this methodology is due to
Luscher~\cite{Luscher:1982uv}, who applied it to
non-gauge, asymptotically free theories in two-dimensions. In that
case, he found that the analytical calculations using semiclassical
methods brought you precisely to values of the observables (ex. mass)  which were rather close to the
final non-perturbative results.  This encouraging result motivated an attempt by various
groups to apply the same ideas to Yang-Mills theory. The study
combined  analytical calculations at small volumes with numerical
results using lattice methods. These numerical methods traditionally constrain  the spatial topology to be
that of a three-dimensional torus, in order to preserve translation
invariance.  

Various groups pursued this path and analyzed the evolution of the 
glueball spectrum as a function of the spatial volume. Some authors
used standard periodic boundary conditions including Luscher
himself~\cite{Luscher:1982ma, Luscher:1983gm}. Perturbation theory and semiclassical methods
encounter great difficulties when formulated on the torus because of
the presence of flat connections~\cite{GonzalezArroyo:1981vw}. These flat connections
depend on continuous parameters giving rise to  zero-modes.
Furthermore, the space of flat connections is an  orbifold having
singular points associated to coinciding eigenvalues, making
collective coordinate methods also rather difficult to apply. Despite this
difficulty Pierre van Baal and collaborators followed this
path~\cite{vanBaal:1986cw, Koller:1987yk, Koller:1987fq}.
In collaboration with  with Chris Korthals Altes and others we also 
followed this strategy but using `t Hooft twisted boundary conditions. 
There were several reasons for this choice, one being that
perturbation theory is much simpler as with certain choices of the
twist tensor there are no continuously connected   flat connections:
hence, no zero-modes.
Furthermore, my previous studies at large N~\cite{GonzalezArroyo:1982hz} showed that twisted
boundary conditions seemed to show that indeed the torus size was
irrelevant at large $N$, realizing the idea put forward by Eguchi 
and Kawai~\cite{Eguchi:1982nm}. 

Our strategy was essentially the following. Start with a small spatial three
dimensional torus of size $l^3$ with twisted boundary conditions and a very large
(essentially  infinite) euclidean temporal direction allowing the extraction of
the mass spectrum by measuring correlations. Some of these masses
become non-zero in perturbation theory and can be computed by standard
methods. Some of our results were collected in a series of
papers~\cite{GonzalezArroyo:1987ycm,Daniel:1989kj, Daniel:1990iz}. 
A particularly simple and attractive case occurs for
SU(2) if we choose a non-zero `t Hooft magnetic flux in all the
planes ($n_{12}\equiv m_3=1$, $n_{23}\equiv m_1=1$ and $n_{31}\equiv m_2=1$). 
This preserves the cubic symmetry under rotations simplifying
the classification of the states according to the corresponding
representations of the group. One of the attractive features of `t
Hooft construction is that the Hilbert space of states becomes a
direct sum of sectors corresponding to different `t Hooft electric
fluxes specified by a 3-vector of integers modulo N (2 for SU(2)).  
At tree level the energies of the ground states of those 
electric flux sectors perpendicular to the 
magnetic charge are lifted (acquire a mass gap). However, the sector 
with electric flux $\vec{e}=(1,1,1)$, parallel to  the magnetic flux,
becomes degenerate with the sector of vanishing flux. We have a
situation with two degenerate ground states at order zero  in
perturbation theory, just as for a double well-potential. With this
structure one can start to compute the energies of the different
states. Obviously these masses have to be proportional to the inverse
length of the torus $1/l$, with  coefficients that appear as a power
series in the coupling constant $g(l)$. For example in the zero
electric flux sector the states appearing above the vacuum can be
called the glueball spectrum. We found that indeed the lowest mass
glueball belongs to the A1 representation corresponding to a scalar.
At a higher mass there are states in the E and $A_2$ representations
which together make up a spin 2 representation of the full rotation group.
This is very reassuring since the ordering matches with that at
infinite volume in contrast with the situation for periodic boundary
conditions.

The energy difference between electric flux sectors measures the
minimal energy carried by this `t Hooft electric flux, which in twisted 
boundary conditions becomes a topological feature. The afore-mentioned
calculations are done in the continuum, but they can also be calculated
numerically by lattice methods. The results agree for small enough
torus sizes as expected from asymptotic freedom. The advantage of the
lattice methodology is that it can also be employed to extract the
energies for much larger torus sizes. In this way one can monitor the
evolution from small sizes for which perturbation theory is a good
approximation to large sizes  which should approach the infinite volume 
case. Since Yang-Mills theory is a gapped theory when the physical
size of the torus $l$ becomes large compared to the corresponding
correlation length the infinite volume features should show up. In
particular, one expects that the results do not depend on the
spatial boundary conditions and that full rotational invariance is recovered.

Concerning the energy of the ground states with non-zero electric
flux $E(\vec{e})$, the expectation is that they tend to infinity linearly with the
torus period with a  coefficient proportional to the string tension:
$E(\vec{e})=\sigma |\vec{e}| l$.
This is Confinement ($|\vec{e}| l$ is the minimal length of a flux
tube carrying this flux). This seems like a great challenge since the
energy corresponding to parallel electric and magnetic flux is zero to
all orders in PT. Understanding how this could happen forces one into
the semiclassical regime. The potential energy has two minima which
differ by the expectation value of a Polyakov loop with winding
numbers equal to the magnetic flux. Indeed, the $\vec{e}=(1,1,1)$ is the
antisymmetric combination of the two minima of the potential and
$\vec{e}=(0,0,0)$ the symmetric one. 
It is a  textbook exercise that in the
semiclassical approximation the gap is generated by the configuration
with least action that interpolates between the two vacua. This is
precisely a fractional instanton of $Q=1/2$. Hence, fractional
instantons do play a role at small torus sizes in generating  a
non-zero energy for the $\vec{e}=(1,1,1)$ sector. That got us
interested in the first place on fractional instantons and led us to show
its existence and its properties. This was part of the thesis of
Margarita Garcia Perez as described earlier~\cite{MGarciaPerez}. We saw that the
fractional instanton tunnels between the two vacua and its
energy profile (action density integrated over space) has a temporal
width which is dictated by the linear size
of the spatial torus (See for example the corresponding profile for
SU(3) in Fig.~5.

\begin{figure}
\includegraphics[width=\linewidth]{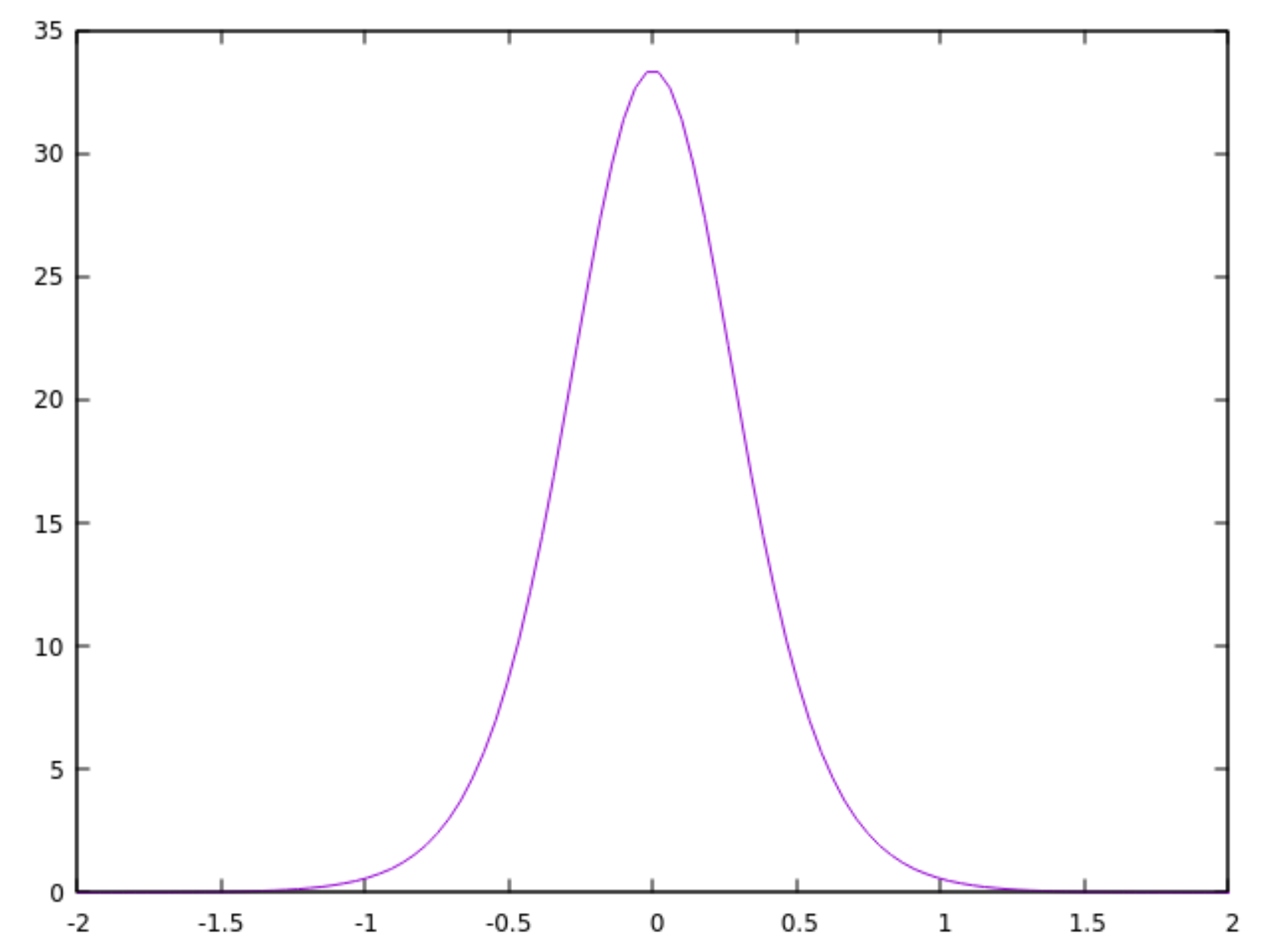}%
\label{fig5}
\caption{This picture shows the Q=1/3 SU(3) fractional instanton energy
profile (action density integrated over space). The figure clearly
show how the solution  tunnels among two minima of the potential. }
\end{figure}

Then,  together with her and with Pablo Martinez we started to apply it to the 
semiclassical understanding of the process interpolating between small and large
distances~\cite{GarciaPerez:1993ab}. A very simple
calculation using the renormalization group shows that the probability
of tunnelling which depends on the free energy of the fractional
instanton grows with the size of the torus. This is in essence the
same phenomenon observed by `t Hooft for ordinary instantons although
the exponent of $l$ is $\sim 8/3$ which is different because the action of fractional
instanton is 1/2 of the one of an instanton. The
semiclassical prediction is  reasonably well reproduced by the numerical
calculation of the mass using the lattice. Furthermore, for small
volumes the lattice configurations can be seen to be composed of a
given  number of fractional instantons (and anti-instantons) and
occasional $Q=1$ instantons. We verified that their distribution in 
euclidean time is  Poisson  and with the probability per unit time as determined by the free
energy. Fig.~6 shows an example of a Monte Carlo generated
configuration after appropriate smoothening which is seen to consist
on several $Q=1/2$ fractional instantons and occasional $Q=1$ ones.
The figure shows both the euclidean time dependence of the energy profile 
as well as the expectation value of the Polyakov line which acts as an
order parameter.  In summary, a beautiful numerical experiment that show how
semiclassical ideas and the renormalization group works.

\begin{figure}
\includegraphics[width=\linewidth]{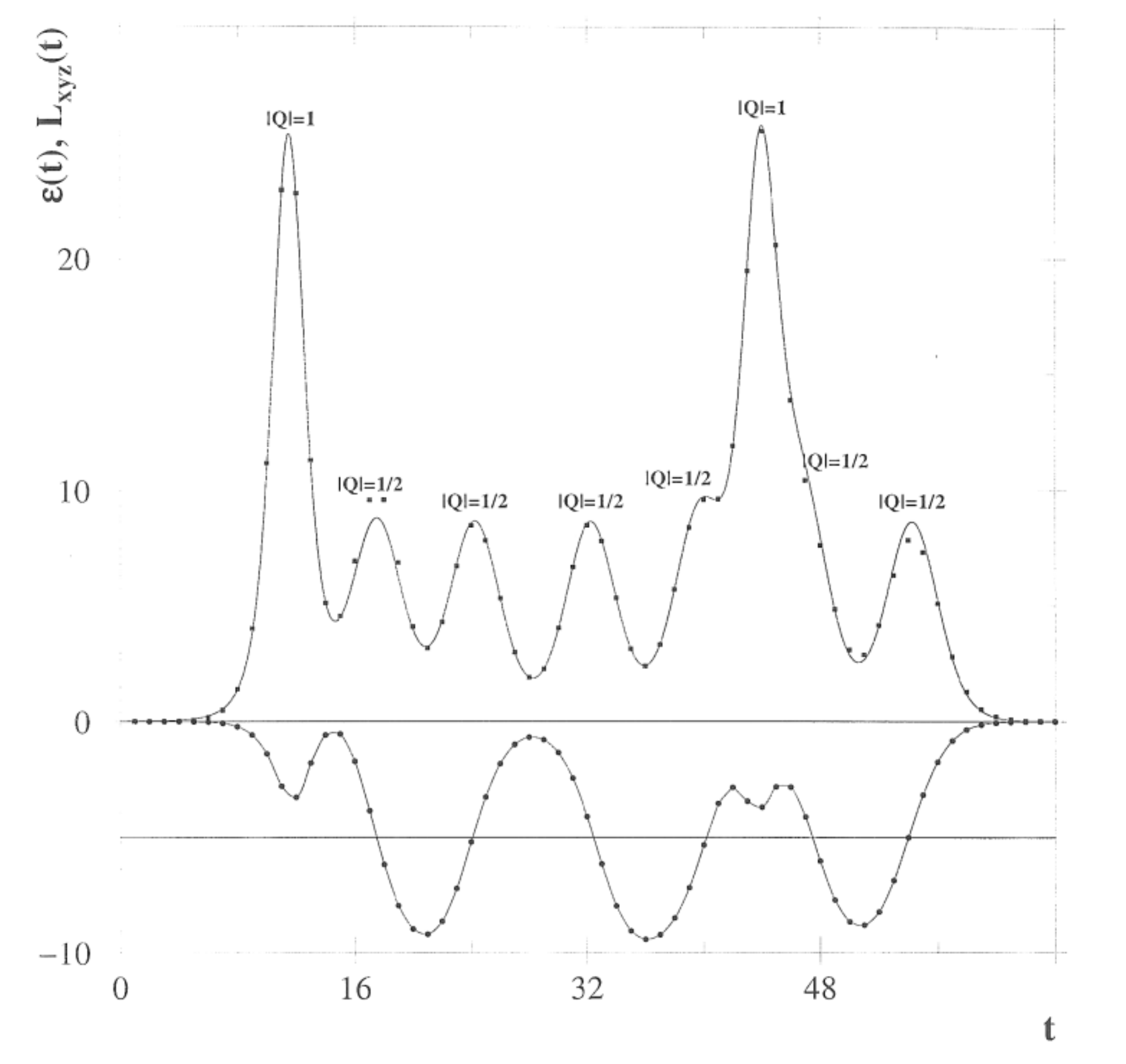}%
\label{fig6}
\caption{This figure depicts the energy profile (upper curve)  of one of the Monte Carlo generated
configurations for SU(2) Yang-Mills with a $T_3\times \R$ topology.
The lower curve shows the behaviour of the trace of the spatial
Polyakov  line which acts as an order parameter. In both cases the
dots are the lattice data after suitable smoothening and the
continuous line are obtained from the classical solution after fitting
the position of the center. }
\end{figure}

As the spatial linear extent in physical units grows, the probability
of generating fractional instantons grows and correspondingly its
density too.  Extrapolating the  semiclassical formula to larger sizes one would predict
that the fractional instantons (which have a finite time-extent) are
now so many as to start overlapping all over the place. This is the
edge of the dilute gas approximation which is on the basis of our
semiclassical computation. For our SU(2) study we estimated this edge
to occur for $l\sim 0.6-0.7\  \mathrm{fermi}$.  The advantage of the numerical calculation
with the lattice is that one can explore also the regions beyond reach
of the semiclassical approximation and all the way to very large
volumes. That is exactly what we did, to discover that once the torus   
is a few fermi in linear size the masses recover the expected behaviour:   rotational
invariance within errors, linear growth of the flux energies and a
string tension matching with the one obtained for large volumes without twist by other
authors.  But the remarkable thing of our result is that the
transition is smooth from the semiclassically dominated region to the
fully large volume confinement region: no transition or cusp and no
level crossings. That is in line with what Luscher found for the
2-dimensional field models. This is summarised in Fig.~7
which shows the ground state energies divided by the length of the
minimal flux tube: $\Sigma(\vec{e})= E(\vec{e})/(|\vec{e}| l)$. 
For large sizes the three different $\Sigma$ should coincide with the
string tension. At small sizes the numerical calculation coincide with
the semiclassical predictions.

\begin{figure}
\includegraphics[width=\linewidth]{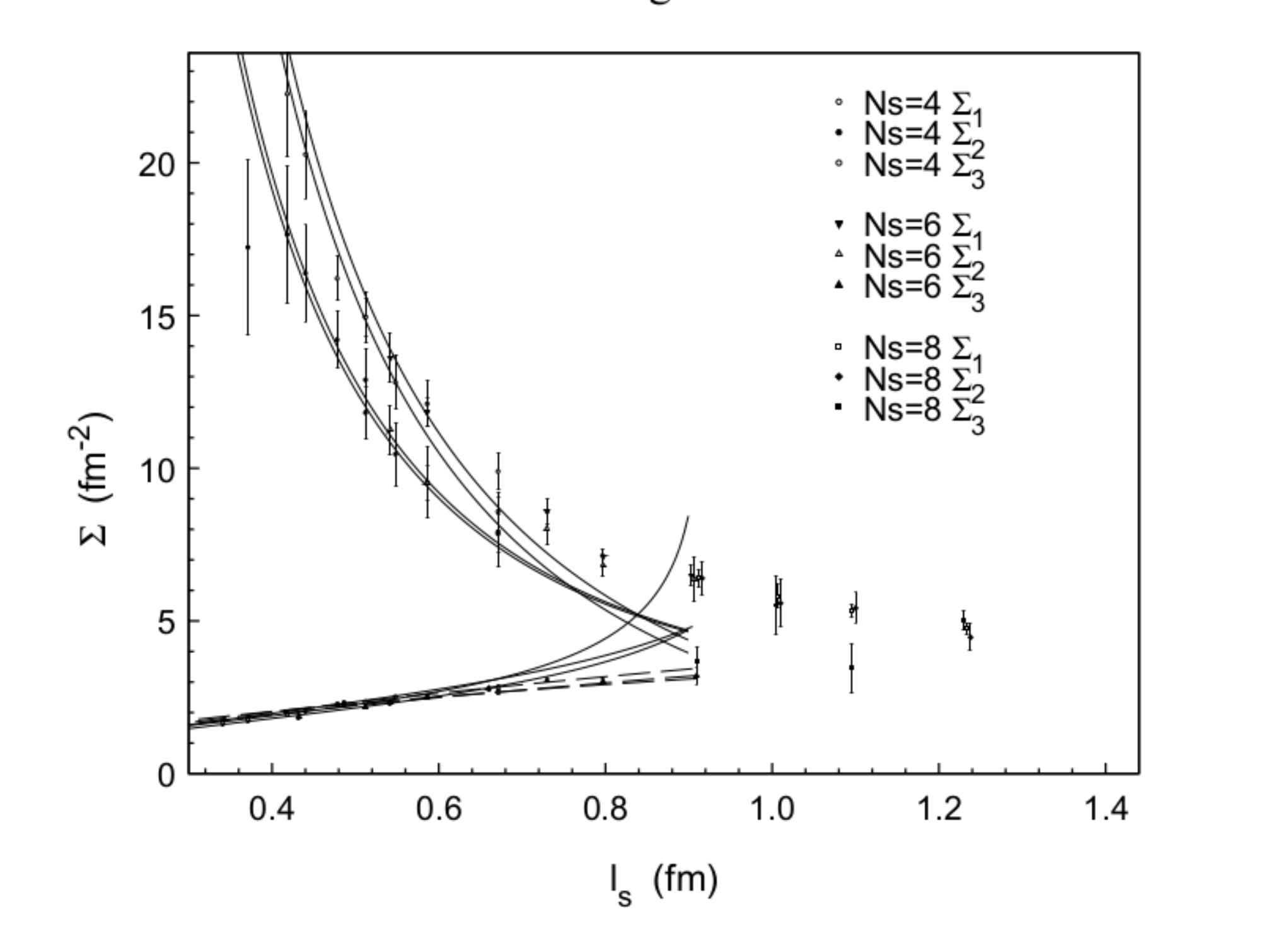}%
\label{fig7}
\caption{This figure~\cite{GonzalezArroyo:1995zy} shows the quantities $\Sigma_1=
E(\vec{e}=(1,0,0))/l_s$, $\Sigma_2= E(\vec{e}=(1,1,0))/(\sqrt{2}l_s)$
and  $\Sigma_3= E(\vec{e}=(1,1,1))/(\sqrt{3}l_s)$ as a function of
the spatial torus size in physical units $l_s$. The curves shown for small 
sizes are the results of the semiclassical calculations. For large $l_s$
Confinement implies that all $\Sigma$ should coincide with the infinite 
volume string tension ($5\ \mathrm{fermi}^{-2}$).}
\end{figure}

How to understand our result? The probability of producing a fractional
instanton is higher the bigger its size. The torus imposes a maximal
size so that for small torus sizes the instanton fills completely the
spatial torus, but is dilute in the temporal direction. However, when the
probability is such that it fills the full temporal size as well it
starts to become competitive to create more instantons of a smaller
size. At one particular point the typical size of the fractional instantons 
is smaller than the torus size and determined by the competition: bigger
instantons or more instantons. This, as mentioned earlier occurs at
$\bar{l}\sim 0.7\ \mathrm{fermi}$. Then as the torus size continues
growing the characteristics of the liquid decouples from that size. In this
way one reaches the infinite volume limit. The just described process 
is smooth and compatible with what we observe. This was then the basis
of our proposed scenario which was  presented in
Ref.~\cite{GonzalezArroyo:1995zy}.

\section{Other properties of the Yang-Mills vacuum}
\label{FinTemp}
Although this was not discussed in our  papers, which focused on
Confinement and Yang-Mills theory in $\R^4$ ($T=0$), our model has the
potential of explaining other non-perturbative properties of the
theory. One is spontaneous chiral symmetry breaking which follows from
the formation of a quark condensate. Here as in the Wilson loop
interpretation,   quark fields appear as non-dynamical sources which
do not back-react on the vacuum. The condensate can be understood
through the Banks-Casher formula~\cite{Banks:1979yr}. The  fractional instantons have an
associated quasi-zero-mode of the Dirac operator. It is of course not
a zero-mode in general since that depends on the total topological
charge which is the result of the cancellation of many fractional
instantons and anti-instantons. Nevertheless, it is these
quasi-zero-modes 
that produce the density entering the Banks-Casher formula. This
mechanism was proposed for ordinary instantons by Diakonov and
Petrov~\cite{Diakonov:1985eg}, but would apply as well for our liquid
model.

The other very important feature of Yang-Mills field theory is the
existence of a finite temperature phase transition. This amounts to
considering fields in euclidean infinite space and periodic in
euclidean time. 
Notice that when the temperature is high the period becomes small and
this limits the size of fractional instantons. This is similar to what
we argued for a small three-dimensional torus. Since the probability
of producing these objects decreases with the size according to the
same renormalization group arguments that we used earlier, the density
is a decreasing function of the temperature. Notice that when we talk
about fractional instantons in this $S_1\times \R^3$ setting we are
actually talking about caloron constituents. How can one then interpret the
Yang-Mills finite temperature phase transition? In our opinion this
would correspond to the percolation transition. For temperatures higher
s than the critical one fractional instantons (calorons) appear in spatial
islands separated from others. At higher temperatures the density of
the islands decreases. On the opposite side, below the transition the islands would
connect into a network of fractional instantons that would recover
more and more the 4D isotropy as we reduce the temperature.

\section{Original  evidence of the FILM model}
4D Yang-Mills field theory is a beautiful theory but it is also a very
difficult one. It is an example of a strongly coupled theory which are
notoriously difficult theories to unravel. Our picture of the vacuum as
a dense media is extremely far from the classical one (flat
connections= pure gauge). How can we test if our model is correct? 

Previously we have argued that the simulations of the theory
using Lattice Gauge theory techniques were reliable in settling the
issue of Confinement and computing the string tension and other
non-perturbative quantities. Hence, the lattice configurations should
contain the answer to our question. The problem is that the theory is
ultraviolet divergent, meaning that local quantities are dominated by
small wavelength  noise. If one displays the action density of a
typical lattice configuration one is not able to distinguish any
pattern.   This happens even if we produce a configuration with
fractional instantons and add quantum fluctuations by hand. It is hard
to visualize the presence of this underlying structures, though there
are still there, as we will argue in a minute.  

\begin{figure}
\includegraphics[width=\linewidth]{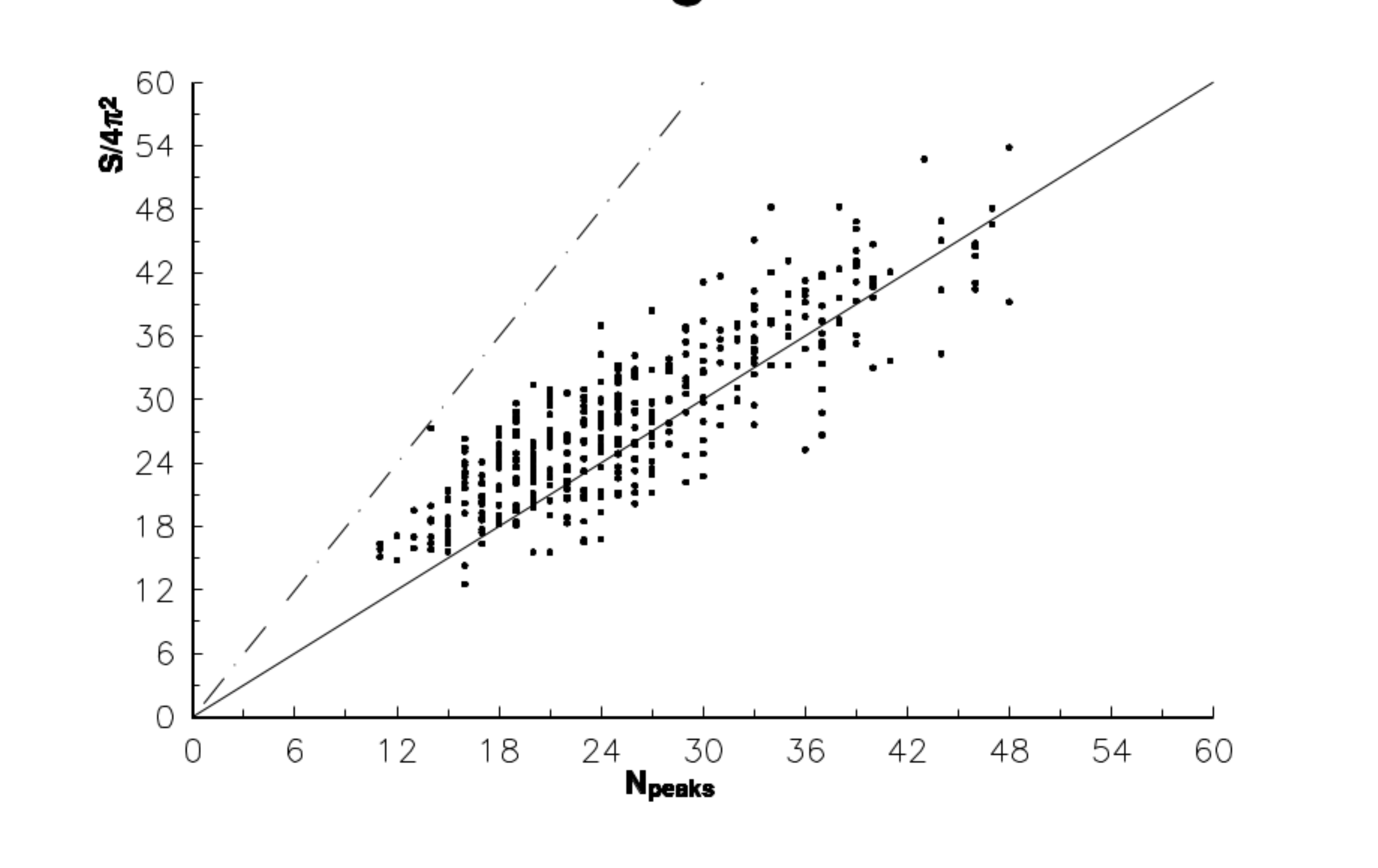}%
\label{fig8}
\caption{This figure~\cite{GonzalezArroyo:1995ex} shows many Monte Carlo
generated configurations for SU(2) Yang-Mills after various levels of smoothening. The
$y$ axis depicts the total action divided by $4\pi^2$ and the $x$-axis
the number of peaks. The lower straight line is the expected one if
the peaks are fractional instantons. The upper one for ordinary
instantons.  }
\end{figure}

To discover any underlying structures one should use some kind of
filtering method to get rid of the high frequency noise without
altering the long-range structure present in the configurations. 
There are a bunch of techniques of this kind available. Most are based 
on local updates that go in the direction of minimizing the action.
At the time of our proposal we employed  the available filtering techniques 
called fuzzying/smearing/cooling ~\cite{Teper:1987wt,
Teper:1987ws, Ape:1987thf, Fernandez:1987ph, Teper:1985rb} applied to
standard SU(2) Yang-Mills Monte Carlo generated lattice configurations. After some
cooling steps we identified local maxima of the action density which
we called {\em peaks}. Are these peaks associated to instantons or
fractional instantons? We measured the average action per peak by
dividing the total action by the number of peaks $N_\mathrm{peaks}$. The result 
is displayed in fig.~\ref{fig8} which appears in
Ref.~\cite{GonzalezArroyo:1995ex}. It is clear that for all
configurations the mean action points towards the identification of
structures with $Q=1/2$  fractional instantons (mean action $4\pi^2$). 
If more cooling steps are applied,  the action decreases but also the number of peaks. 
This is to be expected since the cooling process is bound to
annihilate instantons with neighbouring anti-instantons. However, the
mean action remains approximately the same. Furthermore, we also observed that in
all our configurations included those in
Ref.~\cite{GonzalezArroyo:1996jp}, the string tension and the density
of peaks $\rho$ are correlated in such a way that the dimensionless ratio 
\be
K=\frac{\sigma}{\sqrt{N_\mathrm{peaks}/\mathrm{Volume}}}=\frac{\sigma}{\sqrt{\rho}}
\ee
remains constant for all the configurations with a value 
slightly above $2$. Since the expected density of peaks in the
unflowed liquid is  estimated to be $\rho=1/\bar{l}^4\sim 4\
\mathrm{fermi}^{-4}$ 
the predicted ratio is approximately 2.5 in rough agreement with
measurements. Furthermore, if we assume that the liquid is made out of
fractional instantons and anti-instantons with equal probability that
gives us an estimate of the topological susceptibility of 
\be
\chi=q^2 \rho= \left(\frac{1}{2}\right)^2 \frac{1}{\bar{l}^4} = 1\ 
\mathrm{fermi}^{-4}
\ee
which is also in agreement with measurements. Notice that the value of
$K$ obtained from our randomly produced liquid of fractional
instantons of the same charge is also very
similar~\cite{GonzalezArroyo:1996jp}, validating our argument for
Confinement. Although not included in our old paper it must be said
that according to the percolation hypothesis we also expect that the
critical temperature comes out as $T_c\sim 1/\bar{l}$ which is also in
agreement with lattice determinations. Hence, all the non-perturbative quantities 
are estimated in terms of $\bar{l}$ which can be extracted from the
semiclassical analysis!!

Our early studies also allowed a fairly limited investigation of some
properties of the fluid. In particular we examined  the spatial distribution of
the peaks and found that the preferred Hausdorff dimension was 4. 
This is relevant to determine if the fractional instantons are
preferably arranged as monopole-like 1-dimensional structures or
2-dimensional vortex-like structures. To within the relative precision
of our sample that doesn't seem to be the case.

As mentioned,  all  our early Monte Carlo tests and the quantitative
estimates just given were restricted to  SU(2) Yang-Mills theory because of
computational limitations. In the last section I will comment on the
prospects for exploring larger values of $N$.

\section{Why now?}
The ideas presented in this paper are quite old and published  in the
references that appear in the first section. Furthermore, a summary of
these ideas, shorter but similar to this one, appears in
Ref.~\cite{Perez:2010jx} corresponding to a talk given by me in a workshop 
in Paris in 2009. Hence, one can ask oneself what is the opportunity of 
writing this paper now. I will try to explain it below.

Lately, there has been a renewed interest in these topics and I had
the opportunity of delivering talks at different workshops at ICTS and 
KITP Santa Barbara. My online talk at the ICTS workshop {\em Topological aspects of strong
correlations and gauge theories} in September 2021 was entitled {\em
Fractional Instantons and their dynamical role}. My talk at the KITP
program {\em Confinement, flux tubes and large N}    in February 2022,
was entitled {\em Confinement, fractional instantons, 
large N and all that}. I must also mention  my (general audience)  IMSc Diamond
Jubilee Distinguished Lecture entitled {\em Unveiling the mysteries of
the vacuum of Yang-Mills field theory} delivered on September 2022.

The present manuscript is a somewhat extended version of my talk at
KITP. Apart from enjoying the fantastic  setting and scientific atmosphere in Santa Barbara, 
I had the opportunity of presenting these ideas and having discussions about them with various
participants of which I specially recall those with David Gross,
Mikhail Shifman and Erich Poppitz.  Indeed, it was Poppitz who
suggested me to make a written up version of my talk. 
The main reason is because,  on the bad side,  I could clearly verify that my
work was largely unknown within that audience. Of course, at the time
when the works were done I had the chance to present them in talks in
the Lattice Conferences in Dallas, Bielefeld, Saint Louis, Melbourne
and also in workshops at Trento, Cortona and Moscow 
and in seminars at both sides of the Atlantic. Nonetheless, the
semiclassical ideas that inspire our work were widely disregarded at
that time as applied to Yang-Mills theories. These ideas have however 
become more popular recently in relation to new developments such as
resurgence (see for example Ref.~\cite{Dunne:2014bca,Dunne:2015eaa}). It turns out that works
dealing with Confinement in Yang-Mills theories, fractional instantons,
`t Hooft fluxes, etc are appearing in recent publications in which one 
can read similar  ideas to ours and in which our work is not even cited.
It is then worthwhile to make this manuscript
available to remind  other scientists of our proposal, explain the rationale
behind   and the partial evidence that we obtained  in its support.  

\section{Recent and ongoing progress}
Testing the FILM model  and other models of the Yang-Mills vacuum is a great
challenge for the reasons explained earlier, namely the strongly
coupled nature of the system. We have presented the ideas why we think our
conjecture is very appealing and our early evidence in its favour.
Nonetheless,  as any other conjecture it might be oversimplified. 
Our aim would be that our  description  captures the essence of the
phenomena. But even this  might be completely wrong. 
Although settling the issue conclusively can  take a very long time, there are
prospects for advancing in some the related issues that enter into our description.
At the risk of not giving a complete account,  let me mention a few
aspects  in which there has been recent progress and have the 
potential of providing results in the near future. This includes some
in which I am personally involved.

\begin{enumerate}
\item{\bf Lattice gauge theory studies}\\
Nowadays it is possible to explore larger
values of $N$. Some results have been obtained
recently~\cite{Itou:2018wkm} for SU(3) in which the $T_3\times \R$ 
topology has been used and the corresponding fractional instantons 
emerge clearly.  It would be very interesting to repeat our semiclassical
analysis for $N\ge 3$ to learn something about the distribution of fluxes.
There has been lately some results~\cite{Bribian:2021cmg} for SU(3) which can be
explained by a semiclassical description of the vacuum as a gas-liquid of fractional
instantons but not of ordinary instantons. 

Since we believe that lattice Monte Carlo (MC) configurations are indeed
providing good estimates of the non-perturbative properties of the
theory, we still hope that these MC generated configurations
provide a test of our vacuum model or other. However, as
mentioned earlier the difficulty arises because these configurations
are very noisy and a noise-reduction method has to be applied before
one can discover any underlying long-range structure. The  cooling
methods used in our early studies have been criticized because they
distort the configurations. 
Indeed, even if one uses methods which
control better the effect of lattice
artifacts~\cite{GarciaPerez:1993lic}, one expects the cooling process  to
annihilate neighbouring instanton-anti-instanton pairs. Fortunately, 
the cooling algorithms  are based on local updates and the smoothening proceeds
as a diffusion process from short to large distances. This smearing
radius is under better control with a continuous gradient flow
technique~\cite{Lohmayer:2011si, Luscher:2010iy} which is rather popular nowadays. An alternative is
to use an observable that filters out the short wavelength noise. 
A few years ago we proposed a method~\cite{GonzalezArroyo:2005kh, GarciaPerez:2011tx} based on the Supersymmetric
zero-mode of the adjoint Dirac equation SZM. For smooth solutions of
the equations of motion the density of this mode is proportional to
the action density of the gauge field. However, when adding
short-wavelength noise the amplitude is reduced as opposed to its
effect on the action density. At present I am involved in a
collaboration with Georg Bergner and his PhD student Ivan Soler Calero 
to try to apply this methodology to different MC configurations. 
Apart from extending the   old  $\R\times T_3$  volume dependent study 
 to larger $N$, it is also possible to study    other topologies. 
 For example  a $T_2\times \R^2$ in
which the size of the 2-torus is gradually changed from small values
to large ones. The relevant semiclassical objects are now the
vortex-like arrays of fractional instantons considered in
Refs.~\cite{GonzalezArroyo:1998ez, Montero:1999by, Montero:2000pb}. It is unknown whether one would encounter a phase
transition separating the small  and  large volume regimes. In any
case when the torus size becomes large and one recovers rotational
invariance the final picture would be the same as the one found for
$T_3\times \R$.

Although we have confidence in the lattice approach, it might involve
important computational resources. It would be nice to have a larger
group of researchers working in these aspects. It must be said that
interest in the Yang-Mills vacuum has been present in the lattice
gauge theory community for many years. The effort has
been directed to effective theories based on monopoles or center
vortices which appear after the gauge field configurations have been
appropriately gauge fixed. These pictures are not incompatible with
our proposal. Our description is entirely given in terms of gauge
invariant objects. If one submits the classical fractional instanton
configurations to the aforementioned gauge fixing process one ends up
having monopoles or center vortices located precisely at the location
of the fractional instantons. For example, this is shown explicitly in
Ref.~\cite{Montero:1999by}.
Unfortunately, the gauge-fixing does obscure the microscopic mechanism
underlying Confinement which lies in the  origin of these effective degrees of
freedom.

Finally, although not our main topic, I want to  briefly comment on finite temperature studies.
There are several papers relating
the properties of the theory (mostly QCD) just above the critical temperature with
caloron structures (see for example Ref.~\cite{Larsen:2019sdi}). Notice
that what in these papers are  called instanton-dyons
are just one dimensional arrays of unit fractional instantons. 
It would be very nice to connect these results with the idea of a percolation
transition of fractional instantons.

\item{\bf Self-dual gauge fields in terms of fractional instantons}\\
In constructing our model we argued that the space  of self-dual 
configurations on the torus can  be described as a liquid of unit fractional
instantons.  This is also a conjecture, but a  purely classical one  in which 
powerful mathematics and new ideas might help to settle soon.
The argument implies  that the moduli space of 
self-dual configurations on the torus can be parameterized in terms of
the position of $QN$ unit fractional instantons. We know of one
example in which this has been achieved and that is the case of $Q=1$
calorons~\cite{Lee:1998vu, Lee:1998bb,Kraan:1998sn,Kraan:1998pm}. 
This can serve as a model for an extension to other
cases. There is another case for which we have an analytical solution
and that is the one studied by `t Hooft. However, in that case the
action density is uniform (constant) and talking about the position of
fractional instantons looks bizarre. In our attempt to have
some analytical control over these classical solutions we proposed to
study deformations of `t Hooft solution for geometries that are
slightly deformed with respect the one in which the solution is
self-dual. This was done for
SU(2) in Ref.~\cite{GarciaPerez:2000aiw} and recently extended to
SU(N) in Ref.~\cite{Gonzalez-Arroyo:2019wpu}. 
Even for small deformations a space-time structure emerges and the
notion of fractional instanton position starts to make sense. Our
deformation method allows to set-up a set of equations order by order
in the deformation parameter,  which can be shown to have as many solutions as
indicated by the index theorem. This provides a proof of existence of
these self-dual solutions for geometries differing from that used by `t
Hooft. We believe that the methodology centered around our deformation
method has still to be fully exploited, specially because the SU(N)
case provides several interesting limits. 

Curiously the deformation technique applies in much more general contexts. In
particular, it does apply to the much simpler case of the
2-dimensional abelian Higgs model at the critical value of the
coupling. The Bogolmolny equations are the equivalent of the
self-duality ones. When formulating the theory on a torus, we
encounter a very similar situation. Spatially constant solutions, as
those of `t Hooft,  appear in the so-called  {\em Bradlow limit} as studied by Nick
Manton and others. The name follows the work of the mathematician
Stephen Bradlow on vortex systems on different manifolds~\cite{Bradlow:1990ir}. We studied
the same kind of deformation technique in the thesis work of my
student Alberto Ramos~\cite{GonzalezArroyo:2004xu}. For obvious
reasons we named  the
deformation expansion: {\em Bradlow parameter expansion}. Interestingly
in that case the solution of the Bogomolny equations can be obtained 
iteratively and the solution proven to be unique once the initial step fixes the
position of the zeroes of the Higgs field. These zeroes operate as the
coordinates of the individual unit flux vortices with which higher
flux solutions can be parameterized. Achieving something similar for
fractional instantons would be very nice. Of course the 4D Yang-Mills
case is much harder to handle but as stressed in
Ref.~\cite{Gonzalez-Arroyo:2019wpu} many of
the expressions of the 2D counterpart can be used in that case. 

A possible tool that can be used as well is the Nahm
transform~\cite{Nahm:1979yw, Corrigan:1983sv}. Although originally invented for monopoles, it is in
the study of self-dual solutions on the torus where the method becomes
more beautiful~\cite{Schenk:1986xe, Braam:1988qk, donaldson1990geometry}. It becomes an involution mapping self-dual
configurations onto self-dual configurations on the dual torus. Indeed, it
maps twisted boundary conditions into twisted boundary conditions and 
fractional instantons onto fractional
instantons~\cite{GonzalezArroyo:1998ia}. Because of the mapping to the
dual torus it maps solutions of the type used in our Hamiltonian $T_3\times \R$
analysis to those occuring with calorons producing a unified
picture~\cite{GarciaPerez:1999bc}. Actually, as some torus periods go to infinity the
corresponding Nahm-transform  collapses to a point and produces a dimensional reduction. This is
essentially the idea used by Pierre van Baal to obtain his caloron
solutions.  In a similar spirit the Nahm transform idea illuminates the
origin of the peculiar algebraic conditions entering the ADHM
construction, that can be seen  as the self-duality
equations in a zero-dimensional field theory.

\item{\bf Quantum fluctuations on self-dual solutions}\\
This is also an area in which progress can be achieved in an analytic
way. I already mentioned works that go in this direction mostly for
calorons~\cite{Diakonov:2004jn, Diakonov:2005qa, Gromov:2005hv,
Gromov:2005ij, KorthalsAltes:2014dkx}. In addressing the same type of problems for more
general multi-fractional instanton situations one can also employ the
deformation method mentioned in the previous item. Having analytic
formulas for the gauge fields themselves as a power series in the
deformation parameter, allows to study also quantum fluctuations
analytically. One can then see how the fluctuations might break the
degeneracy of the classical solutions and create a landscape within
the moduli space. This might help in determining the preferred
structure of the liquid since the position of the individual
fractional instantons might prefer some particular arrangement driven
by quantum fluctuations. Of course, this applies only to self-dual
configurations so that it would still leave open what the situation is
when combining fractional instantons and anti-instantons.

\item{\bf Large N}\\
The large N limit of Yang-Mills  field theory~\cite{tHooft:1973alw} is thought to preserve all the  
non-trivial properties of the theory including Confinement. Thus, any
model of the vacuum must pass the test of surviving this limit. This
is non-trivial. For example, it does seem to rule out a proposal based
on ordinary instantons. The challenges continue since one has to find 
in which way one can make compatible the FILM model with  Eguchi-Kawai volume
reduction~\cite{Eguchi:1982nm} as in our implementation of this
idea~\cite{GonzalezArroyo:1982hz}. A nice recent review explaining
progress in large $N$ gauge theory appears in
Ref.~\cite{GarciaPerez:2020gnf}. 

Are there fractional instantons in the large $N$ limit? This was
investigated as part of the thesis of Alvaro Montero and his
subsequent work~\cite{GarciaPerez:1997fq, AMontero, Montero:2000mv}. It seemed that in the large $N$ limit the action
profile  of the $T_3\times\R$ fractional was tending towards step functions. 
Recent results have also studied a different type of fractional
instantons surviving the large $N$ limit~\cite{DasilvaGolan:2022jlm}. It would be very
nice to connect these numerical solutions with the analytic
deformation method of Ref.~\cite{Gonzalez-Arroyo:2019wpu}.

\item{\bf New ideas and methods}\\
As emphasized earlier Yang-Mills field theory in the infrared regime
is a strongly coupled system which makes it a real challenge. 
Since we
are aiming at a microscopic mechanism for Confinement, which is system
dependent,  we do not have powerful tools as Supersymmetry at our
disposal~\cite{Seiberg:1994rs, Seiberg:1994bz}. The same problem applies to
adjoint QCD which includes fermions in the adjoint representation. We
want to mention that this  theory has given rise to a program similar in 
spirit to the one of Luscher but attached to the topology $S_1\times \R^3$. 
The size  of the periodic direction interpolates between the  a semiclassical regime
and the large volume  theory. The smoothness of this transition is
referred as {\em adiabatic continuity} and a mechanism for confinement
for the model in terms of so-called bions emerged from the study (see for example
Ref.~\cite{Unsal:2007jx, Dunne:2016nmc}).

New recent ideas methods~\cite{Gaiotto:2014kfa, Kapustin:2014gua, Gaiotto:2017yup} 
have generated an important activity
also in relation to understanding the presence of `t Hooft flux from a
different perspective. This has led to several papers that have some
relation with our problem~\cite{Unsal:2020yeh, Cox:2021vsa,
Tanizaki:2022ngt}. We hope this methodology could provide new clues to
our problem.

\end{enumerate}

\section*{Acknowledgments}
My gratitude goes first to my former students Margarita García Pérez,
Pablo Martínez, Alvaro Montero,  Carlos Pena, Alberto Ramos and
Alfonso Sastre thanks to whose work and dedication the original ideas 
expressed in this paper were developed. These thanks go very specially 
to Margarita García Pérez from whose ideas, comments and  conversations  I have strongly benefited
over many years of mutual collaboration. In the early times I also benefited
from many conversations with Pierre van Baal who was one of the first to
hear about our model. I already mentioned Erich Poppitz who encouraged
me to write this paper after our long conversations in KITP. I also
enjoyed lengthy conversations and correspondence with Mithat Unsal
over the years since we first met at the Paris meeting in which I made
a previous recollection of the FILM model~\cite{Perez:2010jx}. I  also
thank Sayantan Sharma for the invitation to deliver a IMSC Diamond
Jubilee distinguished lecture and together with the other organizers
for giving me the opportunity to participate in the  ICTS 2021 online workshop
{\em Topological aspects of strong correlations and gauge theories}. 
Very special thanks go to Sergei Dubovsky and the other organizers of
the KITP Santa Barbara program {\em Confinement, fractional instantons and  large
N} for the opportunity to participate and deliver my talk which serves
as a smoking gun for this manuscript. Thanks to KITP and Munger
Residence for the wonderful setting and the funding of the National Science Foundation
under Grant No. NSF PHY-1748958. 

Finally, although my talk was not related to the same topic, I also 
want to acknowledge the Simons Center for  Geometry and Physics at Stony
Brook and the organizers of its program {\em Confronting Large N,
Holography, Integrability and Stringy Models with the Real
World} for inviting me to participate. This allowed me to share conversations related to
the subject matter of this paper with Erich Poppitz, Mithat Unsal, Mohamed Anber, Georg
Bergner, Zohar Komargodski and Jaume Gomis.

A.G-A  work is partially supported by grant PGC2018-094857-B-I00 funded
by “ERDF A way
of making Europe”, by MCIN/AEI/10.13039/501100011033, and by the
Spanish Research Agency
(Agencia Estatal de Investigación) through grants IFT Centro de
Excelencia Severo Ochoa SEV-
2016-0597 and No CEX2020-001007-S, funded by
MCIN/AEI/10.13039/501100011033. We also  acknowledge support from the 
project H2020-MSCAITN-2018-813942 (EuroPLEx) and the EU
Horizon 2020 research and innovation programme, STRONG-2020 project,
under grant agreement No 824093. 

\section*{Disclaimer}
The purpose of this paper was recalling the fractional instanton
liquid model of the vacuum, explaining its characteristics and the
partial evidence that we have in its support. It should not be
understood as a review of Confinement (readers might consult recent
books~\cite{Greensite:2011zz}) or Yang-Mills theory. First, we have stated that our
model applies for 4D  $T=0$ Yang-Mills field theory, so other
interesting theories are left out. Still we cannot rule out that  we have
forgotten to cite  any fully relevant publication that we are not aware of.
We hope to correct any of these in a future revised version.
\bibliography{FILM_PAPER}
\end{document}